\documentclass[letterpaper,11pt]{article}
\usepackage{geometry}
\usepackage{amsmath}
\usepackage{commath}
\usepackage{amsthm}
\usepackage{amssymb}
\usepackage{setspace,mathrsfs}
\usepackage{algorithm,natbib}
\usepackage{algorithmic}
\usepackage{url,amsmath,amsfonts,amssymb}
\usepackage{caption,subcaption,algorithm,graphicx}
\usepackage{listings,enumerate,color,relsize,multirow,rotating,soul}
\usepackage{xr}
\def\Dsc{\mathcal{D}}
\def\Dschat{\widehat{\Dsc}}
\def\Lsc{\mathcal{L}}
\def\bD{\mathbf{D}}

\def\Dscr{\mathscr{D}}
\def\Oscr{\mathscr{O}}
\def\Tscr{\mathscr{T}}

\def\Lsc{\mathcal{L}}
\def\Csc{\mathcal{C}}
\def\Lschat{\widehat{\Lsc}}
\def\argmindum{\mathop{\mbox{argmin}}}
\def\argmin#1{\argmindum_{#1}}

\def\bx{\mathbf{x}}
\def\nminv{n_m^{-1}}
\def\suminm{\sum_{i=1}^{n_m}}
\def\Ninv{N^{-1}}

\def\btheta{\boldsymbol{\theta}}
\def\suponetrans{^{\scriptscriptstyle \sf (1) T}}
\def\suptwotrans{^{\scriptscriptstyle \sf (2) T}}

\def\supbullettrans{^{\scriptscriptstyle \sf (\bullet) T}}
\def\supmtrans{^{\scriptscriptstyle \sf (m) T}}

\def\supMtrans{^{\scriptscriptstyle \sf (M) T}}

\def\supone{^{\scriptscriptstyle \sf (1)}}
\def\suptwo{^{\scriptscriptstyle \sf (2)}}
\def\supthree{^{\scriptscriptstyle \sf (3)}}

\def\supm{^{\scriptscriptstyle \sf (m)}}

\def\supbullet{^{\scriptscriptstyle \sf (\bullet)}}

\def\supM{^{\scriptscriptstyle \sf (M)}}
\def\supth{^{\scriptscriptstyle \sf th}}
\def\bY{\mathbf{Y}}
\def\bX{\mathbf{X}}
\def\Abb{\mathbb{A}}
\def\Xbb{\mathbb{X}}
\def\Tbb{\mathbb{T}}
\def\Zbb{\mathbb{Z}}

\def\Wbb{\mathbb{W}}

\def\subSMA{_{\scriptscriptstyle \sf SMA}}
\def\subadel{_{\scriptscriptstyle \sf SHIR}}
\def\subSfull{_{\Ssc_{\sf full}}}
\def\sublab{_{\scriptscriptstyle \sf L\&B}}
\def\bone{\mathbf{1}}
\def\bzero{\mathbf{0}}

\newtheorem{definition}{Definition}

\newtheorem{cond}{Condition}

\newtheorem{theorem}{Theorem}
\newtheorem{remark}{Remark}

\definecolor{purple}{rgb}{0.84, 0.17, 0.89}
\definecolor{red2}{rgb}{0.7, 0, 0.1}

\usepackage{color}

\newcommand{\bei}{\begin{itemize}}
\newcommand{\eei}{\end{itemize}}
\newcommand{\ben}{\begin{enumerate}}
\newcommand{\een}{\end{enumerate}}
\def\0{\boldsymbol{0}}
\def\u{\boldsymbol{u}}
\def\v{\boldsymbol{v}}

\def\x{\mathbf{x}}

\def\O{\boldsymbol{\Omega}}

\def\Ibb{\mathbb{I}}

\newcommand{\RR}{\mathbb{R}}

\newbox\TempBox \newbox\TempBoxA
\def\ep{\textsf{E}} % the symbol E for expectation used the sans serif letter
\def\trans{^{\scriptscriptstyle \sf T}}

\def\text#1{\mbox{\sf #1}}

\def\bX{\mathbf{X}}
\def\half{\frac{1}{2}}
\def\bmu{\boldsymbol{\mu}}

\def\bbeta{\boldsymbol{\beta}}
\def\boeta{\boldsymbol{\boeta}}
\def\balpha{\boldsymbol{\alpha}}
\def\balphahat{\widehat{\balpha}}
\def\bbetahat{\widehat{\bbeta}}

\def\bbetabreve{\breve{\bbeta}}
\def\Qsc{\mathcal{Q}}
\def\Qschat{\widehat{\Qsc}}

\def\QhatSMA{\Qhat\subSMA}
\def\Qhat{\widehat{Q}}
\def\sublasso{_{\scriptscriptstyle \sf LASSO}}

\def\subdlasso{_{\scriptscriptstyle \sf dLASSO}}
\def\subIPD{_{\scriptscriptstyle \sf IPD}}
\def\bg{\mathbf{g}}
\def\bghat{\widehat{\bg}}
\def\muhat{\widehat{\mu}}
\def\balphahat{\widehat{\balpha}}

\def\half{\frac{1}{2}}
\def\Ssc{\mathcal{S}}
\def\Sschat{\widehat{\Ssc}}

\def\bA{\mathbf{A}}
\def\Vbbbreve{\breve{\Vbb}}

\def\Vbb{\mathbb{V}}
\def\Hbbhat{\widehat{\Hbb}}
\def\Hbb{\mathbb{H}}
\def\bTheta{\boldsymbol{\Theta}}
\def\bThetahat{\widehat{\bTheta}}

\def\Abb{\mathbb{A}}
\def\Xbb{\mathbb{X}}
\def\supeff{^{\scriptscriptstyle \sf eff}}
\def\neff{n\supeff}

\def\Zbb{\mathbb{Z}}

\usepackage{xr}
\begin{document}

%\setstretch{1.3}
\def\spacingset#1{\renewcommand{\baselinestretch}%
{#1}\small\normalsize} \spacingset{1.2}

\title{Individual Data Protected Integrative Regression Analysis of High-dimensional Heterogeneous Data}

\author{Tianxi Cai$^{1,2},$  \  Molei Liu$^{2},$  \ and \ Yin Xia$^{3}$}

\date{}

\footnotetext[1]{Authors are listed in alphabetical order.}

\footnotetext[2]{Department of Biostatistics, Harvard School of Public Health, Harvard University. The research of Tianxi Cai was supported in part by  NIH Grants U54 HG007963 and R21 CA242940.}

\footnotetext[3]{Department of Statistics, School of Management, Fudan University. The research of Yin Xia was supported in part by  NSFC Grants 11771094, 11690013.}

\date{}

\maketitle

% Aggregating Debiased Local Estimator (ADeLE), 
\begin{abstract}
%\vspace{.3in}

Evidence based decision making often relies on meta-analyzing multiple studies, which enables more precise estimation and investigation of generalizability. Integrative analysis of multiple heterogeneous studies is, however, highly challenging in the ultra high dimensional setting. The challenge is even more pronounced when the individual level data cannot be shared across studies, { known as DataSHIELD contraint \citep{wolfson2010datashield}}. Under sparse regression models that are assumed to be similar yet not identical across studies, we propose in this paper a novel integrative estimation procedure for { data-{\bf S}hielding {\bf H}igh-dimensional {\bf I}ntegrative {\bf R}egression (SHIR)}. SHIR protects individual data through summary-statistics-based integrating procedure, accommodates between study heterogeneity in both the covariate distribution and model parameters, and attains consistent variable selection. { Theoretically, SHIR is statistically more efficient than the existing distributed approaches that integrate debiased LASSO estimators from the local sites.} Furthermore, the estimation error incurred by aggregating derived data is negligible compared to the statistical minimax rate and SHIR is shown to be asymptotically equivalent in estimation to the ideal estimator obtained by sharing all data. The finite-sample performance of our method is studied and compared with existing approaches via extensive simulation settings. We further illustrate the utility of SHIR to derive phenotyping algorithms for coronary artery disease using electronic health records data from { multiple chronic disease cohorts}.
\end{abstract}


%%%%%%%%%%%%%%%%%%%%%%%%%%%%%%%%%%%%%%%
%%%%%%%%%%%%%%%%%%%%%%%%%%%%%%%%%%%%%%%

\clearpage
\newpage

\spacingset{1.2} % DON'T change the spacing!
%\newpage

\section{Introduction}
%%%%%%%%%%%%%%%%%%%%%%%%%%%%%%%%%%%%%%%
%%%%%%%%%%%%%%%%%%%%%%%%%%%%%%%%%%%%%%%
\subsection{Background}
Synthesizing information from multiple studies is crucial for evidence based medicine and policy decision making. Meta-analyzing multiple studies allows for more precise estimates and enables investigation of generalizability. In the presence of heterogeneity across studies and high dimensional predictors, such integrative analysis however is highly challenging. %The challenge is even more pronounced when the individual level data cannot be shared across studies due to privacy concerns. 
An example of such integrative analysis is to develop generalizable predictive models using electronic health records (EHR) data 
from different hospitals.
In addition to high dimensional features, EHR data analysis encounters privacy constraints in that individual-level data typically cannot
be shared across local hospital sites, which makes the challenge of integrative analysis even more pronounced. Breach of Privacy arising from data sharing is in fact a growing concern in general for scientific studies. %\citep{erlich2014routes,fecher2015drives} 
Recently, \citet{wolfson2010datashield} proposed a generic individual-information protected integrative analysis framework, named DataSHIELD, that transfers only summary statistics from each distributed local site to the central site for pooled analysis.
%DataSHIELD (Data aggregation through anonymous Summary-statistics from Harmonised Individual levEL Databases)
Conceptually highly valued by  research communities \citep[e.g.]{jones2012datashield,doiron2013data}, the DataSHIELD facilitates important data co-analysis settings where individual-level data meta-analysis (ILMA) is not feasible due to ethical and/or legal restrictions \citep{gaye2014datashield}. { In the low dimensional setting, a number of statistical methods have been developed for distributed analysis that satisfy the DataSHILED constraint \citep[e.g]{chen2006regression,wu2012g,liu2014distributed,lu2015webdisco,huang2015distributed,han2016bootstrap,he2016sparse,zoller2018distributed,duan2019odal,duan2020learning}. Distributed learning methods for high dimensional regression have largely focused on settings without between-study heterogeneity as detailed in Section \ref{sec:intro:relate}. To the best of our knowledge, no existing distributed learning methods can effectively handle both high-dimensionality together with the presence of model heterogeneity across the local sites.}

\subsection{Related Work}\label{sec:intro:relate}
{ In the context of high dimensional regression, several recently proposed distributed inference approaches can be potentially used for the meta-analysis under the DataSHIELD constraint. Specifically, \cite{tang2016method}, \cite{lee2017communication} and \cite{battey2018distributed} proposed distributed inference procedures aggregating the local debiased LASSO estimators \citep{zhang2014confidence,van2014asymptotically,javanmard2014confidence}. By including debiasing procedure in their pipelines, the corresponding estimators can be used for inference directly. \cite{lee2017communication} and \cite{battey2018distributed} proposed to further truncate the aggregated dense debiased estimators to achieve sparsity; see also \cite{maity2019communication}. %{\blue [why do we separate this reference?]} 
Though this debiasing-based  strategy can be extended to fit for our heterogeneous modeling assumption, it still loses statistical efficiency due to the failure to account for the heterogeneity of the information matrices across different sites. In addition, the use of debiasing procedure at local sites incurs additional error for estimation, as detailed in Section 4.4. %In contrasts, our proposal is free of these two issues and thus provides more efficient estimation than the debiasing-based approach. We shall leave relevant details in the next several sections.
}

Besides, \cite{lu2015webdisco} and \cite{li2016supporting} proposed distributed approaches for $\ell_2$-regularized logistic and Cox regression, which rely on iterative communications across the studies. Their methods require sequential communications between local sites and the central machine, which may be time and resource consuming, especially since human effort is often needed to perform the computation and data transfer in many practical settings. { \cite{chen2014split} proposed to estimate high dimensional parameters by first adopting majority voting to select a positive set and then combining local estimation of the coefficients belonging to this set. \cite{wang2014median} proposed to aggregate the local estimators through their median values rather than their mean, shown to be more robust to poor estimation performance of local sites with insufficient sample size \citep{minsker2019distributed}.} More recently, { \cite{wang2017efficient} and \cite{jordan2018communication} presented a communication-efficient surrogate likelihood framework for distributed statistical learning that only transfers the first order summary statistics, i.e. gradient between the local sites and the central site.  \cite{fan2019communication} extended their idea and proposed two iterative distributed optimization algorithms for the general penalized likelihood problems. However, their framework, {as well as others summarized in this paragraph}, is restricted to homogeneous scenarios and cannot be easily extended to the settings with heterogeneous models or covariates. %In summary, there is a paucity of privacy preserving methods that can perform integrative analysis in the presence of heterogeneity. Filling this gap, our method provides a non-trivial generalization of the SMA to the high dimension setting while preserving its nature in accommodating model heterogeneity and the DataSHIELD constraints.

}

%Under a low or moderate dimensional setting, \cite{he2016sparse} proposed a sparse meta-analysis approach (SMA) to integrate information across sites that relies solely on summary statistics and accommodates the heterogeneity of model and covariates. Nevertheless, the SMA method relies on the maximum likelihood estimator (MLE) in each local site and hence is not applicable when $p$ is large relative to $n$. { More recently, \cite{duan2019heterogeneity} extended the above mentioned surrogate likelihood approach \citep{wang2017efficient,jordan2018communication} to the heterogeneous low-dimensional setting using density ratio tilting.} 

\subsection{Our Contribution}
{ In this paper, we fill the methodological gap of high dimensional distributed learning methods that can accommodate cross-study heterogeneity by proposing a novel { data-{\bf S}hielding {\bf H}igh-dimensional {\bf I}ntegrative {\bf R}egression (SHIR)} method under the DataSHIELD constraints. While SHIR can be viewed as analogous to the integrative analysis of debiased local LASSO estimators, it achieves debiasing {\em without} having to perform debiasing for the local estimators.} SHIR solves LASSO problem only once in each local site {\em without} requiring the inverse Hessian matrices or the locally debiased estimators and only needs one turn in communication. %It is thus efficient both in computation and communication. 
Statistically, it serves as the tool for the integrative model estimation and variable selection, in the presence of high dimensionality and heterogeneity in model parameters across sites. In addition, under ultra-high dimensional regime where $p$ can grow exponentially with $n$, SHIR is shown to theoretically achieve asymptotically equivalent performance with the estimator obtained by the ideal individual patient data (IPD) pooled across sites and attain consistent variable selection under some sparsity assumptions. Such properties are not readily available in the existing literature and some novel technical tools are developed for the theoretical verification. { We also show that SHIR is statistically more efficient than the approach based on integrating and truncating locally debiased estimators \citep[e.g.]{lee2017communication,battey2018distributed} through theoretical investigation. Our numerical studies further verify this by comparing our method with the existing approaches. It demonstrates that SHIR enjoys close numerical performance as the ideal IPD estimator and outperforms the other methods.}

%In contrast, the proposed SHIR aims for estimation and variable selection and consequently does not actually require the computation of inverse Hessian matrix as in the aforementioned methods, and is thus computationally more efficient. %Thus, our method requires less assumptions on the structure of the design matrix and is more computationally efficient in that no debiasing is indeed needed when deriving site-specific summary statistics. 
%To compare with their method, we , and as shown in our theoretical and numerical studies, our method has better convergence rate than the debiasing-based strategy and also outperforms it for both the simulation and real example.

\subsection{Outline of the Paper}

The rest of this paper is organized as follows. We introduce the setting in Section \ref{sec:mot} and describe SHIR, our proposed approach in Section \ref{sec:method}. Theoretical properties of SHIR are studied in Section \ref{sec:thm}. We derive the upper bound for its prediction and estimation risks, compare it with the existing approach and show that the errors incurred by aggregating derived data is negligible compared to the statistical minimax rate. When the true model is ultra-sparse, SHIR is shown to be asymptotically equivalent to the IPD estimator and achieves sparsistency. %To the best of our knowledge, in existing distributed learning and inference literature, there is no readily available results on the theoretical equivalence between the integrated estimator and the ideal estimator for  LASSO type high dimensional regression.  
Section \ref{sec:simu} compares the performance of SHIR to existing methods through simulations. We apply SHIR to derive classification models for coronary artery disease (CAD) using EHR data from four different disease cohorts in Section \ref{sec:real}. Section \ref{sec:dis} concludes the paper with a discussion. Technical proofs of the theoretical results are provided in the Supplementary Material.

%{ [The above paragraph was a bit too short in the earlier version, now we emphasize the contributions a bit more.]}

%%%%%%%%%%%%%%%%%%%%%%%%%%%%%%%%%%%%%%%
%%%%%%%%%%%%%%%%%%%%%%%%%%%%%%%%%%%%%%%
\section{Sparse Integrative Analysis}\label{sec:mot}
%%%%%%%%%%%%%%%%%%%%%%%%%%%%%%%%%%%%%%%
%%%%%%%%%%%%%%%%%%%%%%%%%%%%%%%%%%%%%%%

Suppose there are $M$ independent studies and $n_m$ subjects in the $m\supth$ study, for $m = 1, \ldots, M$.
For the $i\supth$ subject in the $m\supth$ study, let $Y_i\supm$ and $\bX_i\supm$ respectively denote the response and the $p$-dimensional covariate vector,  $\bD\supm_i = (Y_i\supm, \bX_i\supmtrans)\trans$, $\bY\supm = (Y_1\supm, \ldots, Y_{n_m}\supm)\trans$, and $\Xbb\supm =(\bX_1\supm ,\bX_2\supm ,\ldots,\bX_{n_m}\supm )\trans$. We assume that the observations in study $m$, $\Dscr\supm = \{\bD_i\supm, i = 1, \ldots, n_m\}$, are independent and identically distributed. Without loss of generality, we assume that $\bX_i\supm$ includes 1 as the first component and $\bX\supm _{i,-1}$ has mean $\mathbf{0}$, where for any vector $\bx=(x_1,x_2,\ldots,x_d)\trans$, $\bx_{-1} = (x_2, \ldots, x_d)\trans$.  We define the population parameters of interest as %\tcomm{you need to deal w/ intercept}We consider a class of commonly used models under which the target population parameter for the $m\supth$ study is defined as
$$
\bbeta_0\supm = \argmin{\bbeta\supm}{\Lsc_m(\bbeta\supm)}, \ \mbox{where}\ \Lsc_m(\bbeta\supm) = \ep\{f(\bbeta\supmtrans\bX_i\supm,Y_i\supm)\}, \ 
\bbeta\supm =(\beta_1\supm ,\beta_2\supm ,\ldots,\beta_p\supm )\trans 
$$
for some specified loss function $f$. %\tcomm{perhaps do not make $\bX_i\supmtrans\bbeta$ since sometimes we have nuisance parameters too} 
Examples of loss function 
include $f(\bbeta\trans\bx,y) = (y - \bbeta\trans\bx)^2$ for linear regression and $f(\bbeta\trans\bx,y) = -y \bbeta\trans\bx + \log\{1+\exp(\bbeta\trans\bx)\}$ for logistic regression. 
We assume that $\bbeta\supm_0$ varies across studies but may share similar support and magnitude, which makes it possible to attain more accurate estimation by jointly estimating the regression models across the studies. We will define such structure more rigorously in Section \ref{sec:int}.

Throughout, for any integer $d$, $[d] = \{1, \ldots, d\}$. For any vector $\bx=(x_1,x_2,\ldots,x_d)\trans\in \RR^d$ and index set $\Ssc = \{j_1,\ldots,j_k: j_1<\cdots<j_k\} \subseteq [d]$, 
$\bx_{\Ssc} = (x_{j_1}, \ldots, x_{j_k})\trans$, $\|\bx\|_q$ denotes the $\ell_q$ norm of $\bx$ and $\|\x\|_{\infty}=\max_{j\in[d]}|x_j|$. 
For any matrix $\Abb = [\bA_1, \ldots, \bA_d] \in \RR^{n \times d}$ and index set $\Ssc_1, \Ssc_2 \subseteq [d]$,
let $\Abb_{j\bullet}$ and $\Abb_{\bullet j}$ respectively denote the $j\supth$ row and column of $\mathbb{A}$, $\Abb_{\Ssc_1\Ssc_2}$ denote the submatrix corresponding to rows in $\Ssc_1$ and columns in $\Ssc_2$, $\Abb_{\bullet\Ssc} = [\Abb_{\bullet j_1}, \ldots, \Abb_{\bullet j_k}]$, $\|\mathbb{A}\|_2:=\left[\Lambda_{\max}(\mathbb{A}\trans\mathbb{A})\right]^{\half}$, $\|\mathbb{A}\|_{\infty}=\max_j\|\Abb_{j\bullet}\|_1$. 
%. Define $\v_{\mathcal{S}}=(\v\suponetrans_{\mathcal{S}},...,\v\supKtrans_{\mathcal{S}})\trans$ for ${\mathcal{S}}\subset[d]$, {where $\v\supktrans_{\mathcal{S}}$ represents the subsets of $\v\supktrans$ with indices belong to the set ${\mathcal{S}}$, for each $k=1,...,K$.}
%%
Let %$\Xbb={\sf diag}\{\Xbb\supone, \ldots,\Xbb\supM\}$, 
$\bbeta_j=(\beta\supone_j,\ldots,\beta\supM_j)\trans$, $\bbeta\supbullet=(\bbeta\suponetrans,\ldots,\bbeta\supMtrans)\trans$, 
and let $\bbeta_{0j}$, $\bbeta_0\supm$ and $\bbeta_0\supbullet=(\bbeta_0\suponetrans,\bbeta_0\suptwotrans,\ldots,\bbeta_0\supMtrans)\trans$ respectively denote the true values of $\bbeta_j$, $\bbeta\supm$ and $\bbeta\supbullet$.

%%%%%%%%%%%%%%%%%%%%%%%%%%%%%%%%%%%%%%%
\subsection{Integrative Analysis of $M$ Studies}\label{sec:int}
%%%%%%%%%%%%%%%%%%%%%%%%%%%%%%%%%%%%%%%
To estimate $\bbeta\supbullet_0$ based on $\Dscr\supbullet = \{\Dscr\supone, \ldots, \Dscr\supM\}$, consider the empirical global loss function
\[
\Lschat(\bbeta\supbullet)=\Ninv \sum_{m=1}^Mn_m\Lschat_m(\bbeta\supm),  
\]
where $N=\sum_{m=1}^Mn_m$ and
\[
\Lschat_m(\bbeta\supm)=\nminv\suminm f(\bbeta\supmtrans\bX_i\supm,Y\supm_i), \quad m = 1, \ldots, M.
\]
Minimizing $\Lschat(\bbeta\supbullet)$ is obviously equivalent to estimating $\bbeta\supm$ using $\Dscr\supm$ only. 
To improve the estimation of $\bbeta_0\supbullet$ by synthesizing information from $\Dscr\supbullet$ and overcome the high dimensionality, we assume 
that $\bbeta_0\supbullet$ is sparse and employ penalized loss functions, 
$\Lschat(\bbeta\supbullet) + \lambda\rho(\bbeta\supbullet)$, with the penalty function $\rho(\cdot)$ designed to leverage prior information on $\bbeta_0\supbullet$.

Under a prior assumption that $\bbeta_{0,-1}\supone, \ldots, \bbeta_{0,-1}\supM$, { i.e. the true coefficients excluding intercepts}, share the same support, we may choose $\rho(\cdot)$ as the group LASSO penalty \citep{yuan2006model} $\rho_1(\bbeta\supbullet)=\sum_{j=2}^p\|\bbeta_j\|_2$.  Alternatively, following typical meta-analysis, we decompose $\beta_j\supm$ as $\beta_j\supm=\mu_j + \alpha_j\supm$ with $\balpha_{j}=(\alpha\supone_j,\ldots,\alpha\supM_j)\trans$ and $\bone_{M\times 1}\trans \balpha_j=0$ for identifiability. Similarly as \cite{cheng2015identification}, 
we may impose a mixture of LASSO and group LASSO penalty : %\lcomm{need to comment on why different from Cheng et al}
\begin{equation}
\rho_2(\bbeta\supbullet)=\sum_{j=2}^p|\mu_j|+{\lambda_g}\sum_{j=2}^p\|\balpha_{j}\|_2, \quad 
\label{equ:pen}
\end{equation}
where $\lambda_g \ge 0$ is a tuning parameter. For each $j$, $\mu_j$ represents average effect of the covariate $X_j$ and $\balpha_j$ captures the between study heterogeneity of the effects. This penalty accommodates three type of covariates: (i) homogeneous effect with $\mu_j\neq0$ and $\balpha_j=\mathbf{0}$; (ii) heterogeneous effect with $\balpha_j\neq\mathbf{0}$; (iii) null effect with $\mu_j=0$ and $\balpha_j=\mathbf{0}$.   The penalty $\rho_2$ differs slightly from that of \cite{cheng2015identification} where $\|\balpha_{j,-1}\|_2$ was used instead of $\|\balpha_{j}\|_2$. This modified penalty leads to two main advantages: (1) the estimator is invariant to the permutation of the indices of the $M$ studies; and (2) it yields better theoretical estimation error bounds for the heterogeneous effects detailed as in the proofs. Other penalties that attain similar properties include the hierarchical LASSO penalty \citep{zhou2010group} $\rho_3(\bbeta\supbullet)=\sum_{j=2}^p\|\bbeta_j\|_1^{1/2}$.
Although our proposed SHIR framework can accommodate a variety of sparsity inducing penalties, our theoretical derivations and implementation focus on the mixture penalty $\rho(\cdot) = \rho_2(\cdot)$ throughout but we comment on extensions of SHIR to incorporating alternative penalty functions in Section \ref{sec:dis}. 
%Another alternative penalty function is the hierarchical penalty $\rho_3(\bbeta\supbullet)=\sum_{j=2}^p|\mu_j|+{\lambda_g}\sum_{j=2}^p\|\balpha_{j}\|_1$  \citep{zhou2010group}, which also allows for both shared effects $\{\mu_j\}$ and heterogeneous effects $\{\balpha_j\}$. 

%%

With a given penalty function $\rho(\cdot)$, we may form a penalized loss $\Qschat(\bbeta\supbullet)=\Lschat(\bbeta\supbullet)+\lambda\rho(\bbeta\supbullet)$  and an idealized {\em IPD estimator} for $\bbeta_0\supbullet$ can be obtained as
\begin{equation}
\bbetahat\supbullet\subIPD = \argmin{\bbeta\supbullet}{\Qschat(\bbeta\supbullet)},
\label{equ:glob}
\end{equation}
with some tuning parameter $\lambda \ge 0$.
%{ Athough the IPD estimator is not feasible in practice, it is a helpful benchmark for both theoretical and numerical analysis to assess efficiency loss, if any, due to DataShield constraint, as detailed in Sections \ref{sec:equiv} and \ref{sec:simu}.}
%%
%%%%%%%%%%%%%%%%%%%%%%%%%%%%%%%%%%%%%%%
%\subsection{Sparse Meta-analysis (SMA) Approach}\label{sec:sma}
%%%%%%%%%%%%%%%%%%%%%%%%%%%%%%%%%%%%%%%
However, the IPD estimator is not feasible under the DataSHIELD constraint. Our goal is to construct an alternative estimator that asymptotically attains the same efficiency as $\bbetahat\supbullet\subIPD$ but only requires sharing summary data. When  $p$ is not large, the sparse meta analysis (SMA) approach by \cite{he2016sparse} achieves this goal via a second order Taylor expansion and estimates $\bbeta\supbullet$ as $\bbetahat\supbullet\subSMA = \argmin{\bbeta\supbullet}{\QhatSMA(\bbeta\supbullet)}$, where
\begin{equation}
\QhatSMA(\bbeta\supbullet)=\Ninv \sum_{m=1}^M(\bbeta\supm -\bbetabreve\supm )\trans \Vbbbreve_m^{-1}(\bbeta\supm -\bbetabreve\supm )+\lambda\rho_3(\bbeta\supbullet),
\label{equ:1}
\end{equation}
%\tcomm{did they choose a specific $\rho$?} \tcomm{They used hierarchical lasso, do we need to introduce it in specific?}
$\bbetabreve\supm =\argmin{\bbeta\supm }{\Lschat_m(\bbeta\supm)}$ and  $\Vbbbreve_m=\{n_m^{-1}\nabla^2\Lschat_m(\bbetabreve\supm )\}^{-1}.$
When $\bbeta\supm$ is the same across sites and $\lambda=0$,  $\bbetahat\supbullet\subSMA$ is the inverse variance weighted estimator  \citep{lin2010relative}. Although the SMA attains oracle property for a relatively small $p$,  it fails when $p$ is large due to the failure of $\bbetabreve\supm$.

%The Hessian matrix is singular in the high-dimensional setting, and $\bThetahat\supm$ is used to approximate the inverse of $\Hbbhat_m$ so that $\bThetahat\supm \Hbbhat_m\approx I$. Let the true coefficients be $\bbeta_0\supbullet$, \citep{van2014asymptotically} showed that $\bbetahat\subdlasso\supm$ can be decomposed as the sum of a variance term $U\supm$ and a bias term $\Delta\supm$
%\[\sqrt{N}(\bbetahat\subdlasso\supm-\bbeta_0\supbullet)=U\supm +\Delta\supm .\]
%Under certain conditions, the bias term $\Delta\supm =o(1)$ and the variance term satisfies $U\supm \sim {\sf N}(\mathbf{0},\bThetahat\supm \nabla^2\Lschat_m(\widehat{\beta}_l\supm )\bThetahat\supm ).$ Note that by $\bThetahat\supm \Hbbhat_m\approx I$, the variance of $\sqrt{N}(\bbetahat\subdlasso\supm-\beta_0\supm )$ can be approximately expressed as $\bThetahat\supm \nabla^2\Lschat_m(\widehat{\beta}_l\supm )\bThetahat\supm \approx\bThetahat\supm .$

%%%%%%%%%%%%%%%%%%%%%%%%%%%%%%%%%%%%%%%
%%%%%%%%%%%%%%%%%%%%%%%%%%%%%%%%%%%%%%%
\section{data-Shielding High Dimensional Integrative Regression (SHIR)}\label{sec:method}
%%%%%%%%%%%%%%%%%%%%%%%%%%%%%%%%%%%%%%%
%%%%%%%%%%%%%%%%%%%%%%%%%%%%%%%%%%%%%%%

%%%%%%%%%%%%%%%%%%%%%%%%%%%%%%%%%%%%%%%
\subsection{SHIR Method}\label{sec:SHIR }
%%%%%%%%%%%%%%%%%%%%%%%%%%%%%%%%%%%%%%%
In the high dimensional setting, one may overcome the limitation of the SMA approach by replacing $\bbetabreve\supm$ with the regularized LASSO estimator, 
\begin{equation}
\bbetahat\sublasso\supm = \argmin{\bbeta\supm}{\Lschat_m(\bbeta\supm) + \lambda_m \|\bbeta_{-1}\supm\|_1}
\label{equ:loc}
\end{equation}
However, aggregating $\{\bbetahat\sublasso\supm, m \in [M]\}$ is problematic with large $p$ due to their inherent biases. To overcome the bias issue, 
we build the SHIR  method motivated by SMA and the debiasing approach for LASSO  \citep{van2014asymptotically} yet achieve debiasing {\em without} having to perform debiasing for $M$ local estimators. { Specifically,  we propose the SHIR estimator for $\bbeta_0\supbullet$ as
$\bbetahat\subadel\supbullet = \argmin{\bbeta\supbullet}{\Qhat\subadel(\bbeta\supbullet)}$, where
\begin{equation}
\Qhat\subadel(\bbeta\supbullet)=\Ninv \sum_{m=1}^Mn_m\left\{\bbeta\supmtrans \Hbbhat_m\bbeta\supm -2\bbeta\supmtrans \bghat_m\right\}+\lambda\rho(\bbeta\supbullet),
\label{equ:loss}
\end{equation}
$\Hbbhat_m=\nabla^2\Lschat_m(\widehat\bbeta\supm\sublasso)$ is an estimate of the Hessian matrix and $\bghat_m=\Hbbhat_m\bbetahat\sublasso\supm - \nabla\Lschat_m(\bbetahat\sublasso\supm)$. %derived with local data $\Dscr\supm$.
Our SHIR estimator $\bbetahat\subadel\supbullet$ satisfy the DataSHIELD constraint as $\Qhat\subadel(\bbeta\supbullet)$ depends on $\Dscr\supm$ only through summary statistics $\Dschat_m = \{n_m, \Hbbhat_m, \bghat_m\}$, which can be obtained within the $m\supth$ study, and requires only one round of data transfer from local sites to the central node.}

%{ We consider two choices for $\Hbbhat_m$ and $\bghat_m$. First, we construct them empirically as $\Hbbhat_m=-\nabla^2\Lsc_m(\bbetahat\sublasso\supm)$ and $\bghat_m = \Hbbhat_m\bbetahat\sublasso\supm - \nabla\Lsc_m(\bbetahat\sublasso\supm)$. Then we also consider cross-fitted estimators by splitting $\Dscr\supm$ into $K$ folds $\{\Dscr_1\supm, ..., \Dscr_K\supm\}$. We fit (\ref{equ:loc}) with data $\Dscr\supm\setminus \Dscr\supm_k$ to obtain $\bbetahat\sublasso\supmnk$ and obtain the empirical Hessian and gradient using $\Dscr_k\supm$ evaluated at $\bbetahat\sublasso\supmnk$ for each $k$; and then average over the $K$ folds to obtain the final estimators $\Hbbhat_m$ and $\bghat_m$. The cross-fitting strategy enables us to relax our theoretical assumptions as reflected in Appendix \ref{sec:sup:comp} of the Supplement, similar to those shown in recent work \citep[e.g.]{liu2020integrative,ma2020global}.} 
%%

With $\{\Hbbhat_m, \bghat_m, m = 1, ..., M\}$, we may implement the SHIR procedure using coordinate descent algorithms \citep{friedman2010note} along with reparameterization. Specifically, let $\bmu=(\mu_1,\ldots,\mu_p)\trans$, $\balpha\supbullet=(\balpha\suponetrans, \ldots, \balpha\supMtrans)\trans$, and $\balpha\supbullet_{-1}=(\balpha\suponetrans_{-1}, \ldots, \balpha\supMtrans_{-1})\trans$. Also let $\bmu_0$ and $\balpha\supbullet_0$ be the true value of $\bmu$ and $\balpha\supbullet$, respectively. Let 
$$
\Qhat\subadel(\bmu,\balpha\supbullet)=\Lschat\subadel(\bmu,\balpha\supbullet)+\lambda\rho_2(\bmu,\balpha\supbullet;\lambda_g),$$ 
where $\rho_2(\bmu,\balpha\supbullet;\lambda_g)=\|\bmu_{-1}\|_1+ \lambda_g \|\balpha\supbullet_{-1}\|_{2,1}$, $\|\balpha\supbullet_{-1}\|_{2,1} = \sum_{j=2}^p\|\balpha_{j}\|_2$ and
\[
\Lschat\subadel(\bmu,\balpha\supbullet)=\Ninv \sum_{m=1}^M n_m\left\{(\bmu\trans+\balpha\supmtrans)\Hbbhat_m(\bmu+\balpha\supm) -2\bghat_m\trans(\bmu+\balpha\supm)\right\}.
\]
Then the optimization problem in (\ref{equ:loss}) can be reparameterized and represented as:
\begin{equation*}
\begin{split}
(\widehat{\bmu}\subadel,\widehat{\balpha}\supbullet\subadel)=\argmin{(\bmu,\balpha\supbullet)}~\Qhat\subadel(\bmu,\balpha\supbullet),~~\mbox{s.t.}~\bone_{M\times 1}\trans \balpha_j=0,~j\in[p],
\end{split}
\end{equation*}
and $\widehat{\bbeta}\subadel$ is obtained with the transformation: $\beta_j\supm=\mu_j + \alpha_j\supm$ for every $j\in[p]$.

{
\begin{remark}
The first term in $\Qhat\subadel(\bbeta\supbullet)$ is essentially the second order Taylor expansion of $\Lschat(\bbeta\supbullet)$ at the local LASSO estimators $\bbetahat\sublasso\supbullet$.  The SHIR method can also be viewed as approximately aggregating local debiased LASSO estimators without actually carrying out the 
standard debiasing process. To see this, let
$$
\Qhat\subdlasso(\bbeta\supbullet) = \Ninv \sum_{m=1}^M n_m(\bbeta\supm -\bbetahat\subdlasso\supm)\trans \Hbbhat_m(\bbeta\supm -\bbetahat\subdlasso\supm)+\lambda\rho(\bbeta\supbullet) 
$$
where $\bbetahat\supm\subdlasso$ is the debiased LASSO estimator for the $m\supth$ study with 
\begin{equation}
\bbetahat\supm\subdlasso=\bbetahat\sublasso\supm-\bThetahat_m \nabla \Lschat_m(\bbetahat\sublasso\supm), \quad \mbox{for $m = 1, \ldots, M$},
\label{equ:debias}
\end{equation}
and $\bThetahat_m$ is a regularized inverse of $\Hbbhat_m$. We may write
\begin{equation*}
\begin{split}
\Qhat\subdlasso(\bbeta\supbullet)  {= } \Ninv \sum_{m=1}^M&\left\{n_m\left[\bbeta\supmtrans \Hbbhat_m\bbeta\supm -2\bbeta\supmtrans \Hbbhat_m\bbetahat\subdlasso\supm\right]+C_m \right\} + \lambda\rho(\bbeta\supbullet)\\
\approx \Ninv \sum_{m=1}^M&\left\{n_m\left[\bbeta\supmtrans \Hbbhat_m\bbeta\supm -2\bbeta\supmtrans \bghat_m\right]+C_m \right\}
+\lambda\rho(\bbeta\supbullet) = \Qhat\subadel(\bbeta\supbullet) + C_m,
\end{split}
\end{equation*}
where we use $\bThetahat_m \Hbbhat_m\approx \Ibb$ in the above approximation and the term
\[
C_m=\left\{\Hbbhat_m\bbetahat\sublasso\supm- \Hbbhat_m\bThetahat_m\nabla \Lschat_m(\bbetahat\sublasso\supm)\right\}\trans\left\{\bbetahat\sublasso\supm-\bThetahat_m \nabla \Lschat_m(\bbetahat\sublasso\supm)\right\}
\] 
does not depend on $\bbeta\supbullet$. We only use $\bThetahat_m \Hbbhat_m\approx \Ibb$ heuristically above to show a connection between our SHIR estimator and the debiased LASSO, but the validity and asymptotic properties of the SHIR estimator do not require obtaining any $\bThetahat_m$ or establishing a theoretical guarantee for $\bThetahat_m \Hbbhat_m$ being sufficiently close to $\Ibb$. %which is independent of the target parameter $\bbeta\supbullet$ and thus can be dropped in the objective function in \eqref{equ:loss}. 
\end{remark}

\begin{remark}
Compared with existing debiasing-type methods \citep{lee2017communication,battey2018distributed}, the SHIR  is also computationally and statistically efficient as it is performed without relying on the debiased statistics \eqref{equ:debias} and achieves debiasing without calculating $\bThetahat_m$, which can only be estimated well under strong conditions \citep{van2014asymptotically,jankova2016confidence}. 
\end{remark}
}

%%%%%%%%%%%%%%%%%%%%%%%%%%%%%%%%%%%%%%%
\subsection{Tuning Parameter Selection}\label{sec:tun}
%%%%%%%%%%%%%%%%%%%%%%%%%%%%%%%%%%%%%%%
The implementation of SHIR  requires selection of three sets of tuning parameters, $\{\lambda_m, m \in [M]\}$, $\lambda$ and $\lambda_g$.  
We select $\{\lambda_m, m \in [M]\}$ for the LASSO problem locally via the standard $K$-fold cross validation (CV). Selecting  $\lambda$ and $\lambda_g$ needs to balance the trade-off between the model's degrees of freedom, denoted by $\mbox{DF}(\lambda,\lambda_g)$, and the quadratic loss in $\Qhat\subadel(\bbeta\supbullet)$. It is not feasible to tune $\lambda$ and $\lambda_g$ via the CV since individual-level data are not available in the central site. We propose to select $\lambda$ and $\lambda_g$ as the
minimizer of the generalized information criterion (GIC) \citep{wang2007unified,zhang2010regularization}, defined as
\[
{\rm GIC}({\lambda,\lambda_g})={\rm Deviance}(\lambda,\lambda_g)+\gamma_N\mbox{DF}(\lambda,\lambda_g),
\]
%{ [May be asked whether the $(\lambda,\lambda_g)$ selected by GIC satisfy the conditions in the theorems. Can add a bit discussion on this or later in the revision if asked.]}
where $\gamma_N$ is some pre-specified scaling parameter and
\[
{\rm Deviance}(\lambda,\lambda_g)=\Ninv \sum_{m=1}^Mn_m\left\{\bbetahat\subadel\supmtrans(\lambda,\lambda_g)  \Hbbhat_m\bbetahat\subadel\supm(\lambda,\lambda_g) -2\bghat_m\trans \bbetahat\subadel\supm(\lambda,\lambda_g) \right\} .
\]
Following \cite{zhang2010regularization} and \cite{vaiter2012degrees}, we define $\mbox{DF}(\lambda,\lambda_g)$ as the trace of
\[
\left[\partial^2_{\Sschat_{\mu}, \Sschat_{\alpha}}\Qhat\subadel(\widehat{\bmu}\subadel,\widehat{\balpha}\supbullet\subadel)\right]^{-1}\left[\partial^2_{\Sschat_{\mu}, \Sschat_{\alpha}}\Lschat\subadel(\widehat{\bmu}\subadel,\widehat{\balpha}\supbullet\subadel)\right],
\]
%\begin{align*}
%\left[2\Ninv\Wbb_{\Sschat_{\mu},\Sschat_{\alpha}} (\bbetahat\sublasso\supbullet)\trans\Wbb_{\Sschat_{\mu},\Sschat_{\alpha}}(\bbetahat\sublasso\supbullet)+\partial^2_{\Sschat_{\mu}, \Sschat_{\alpha}}\rho_2(\bmu,\balpha\supbullet;\lambda_g)\right]^{-1}\left[2\Ninv\Wbb_{\Sschat_{\mu},\Sschat_{\alpha}}(\bbetahat\sublasso\supbullet)\trans\Wbb_{\Sschat_{\mu},\Sschat_{\alpha}}(\bbetahat\sublasso\supbullet)\right]
%\end{align*}
where $\Sschat_{\mu}=\{j:\muhat_{{\scriptscriptstyle \sf SHIR , }j}(\lambda,\lambda_g)\neq 0\}$, $\Sschat_{\alpha}=\{j:\|\widehat{\balpha}_{{\scriptscriptstyle \sf SHIR , }j}(\lambda,\lambda_g)\|_2\neq0\}$, {the operator $\partial^2_{\Sschat_{\mu}, \Sschat_{\alpha}}$ is defined as the second order partial derivative with respect to $(\bmu_{\Sschat_{\mu}}\trans,\balpha\suptwotrans_{\Sschat_{\alpha}},\ldots,\balpha\supMtrans_{\Sschat_{\alpha}})\trans,$ 
after plugging $\balpha\supone=-\sum_{m=2}^M\balpha\supm$ into $\Qhat\subadel(\bmu,\balpha\supbullet)$ or $\Lschat\subadel(\bmu,\balpha\supbullet)$.} %\tcomm{Molei: i made some simplifications, plz double check}

\begin{remark}
As discussed in \cite{kim2012consistent}, $\gamma_N$ can be chosen depending on the goal with commonly choices including $\gamma_N=2/N$ for AIC \citep{akaike1974new}, $\gamma_N=\log N /N$ for BIC \citep{bhat2010derivation}, $\gamma_N=\log\log p\log N/N$ for modified BIC \citep{wang2009shrinkage} and  $\gamma_N=2\log p/N$ for RIC \citep{foster1994risk}. We used the BIC with $\gamma_N=\log N /N$ in our numerical studies.
\end{remark}
\begin{remark}
{ For linear models, it has been shown that the proper choice of $\gamma_N$ guarantees GIC's model selection consistency under various divergence rates of the dimension $p$ \citep{kim2012consistent}.} For example, for fixed $p$, GIC is consistent if $N\gamma_N\rightarrow\infty$ and $\gamma_N\rightarrow0$.
When $p$ diverges in polynomial rate $N^{\xi}$, then GIC is consistent provided that $\gamma_N=\log N/N$ (BIC) if $0<\xi<1/2$; $\gamma_N=\log\log p\log N/N$ (modified BIC) if $0<\xi<1$. When $p$ diverges in exponential rate $O(\exp(\kappa N^{\xi}))$ with $0<\nu<\xi$, GIC is consistent as $\gamma_N=N^{\nu-1}$. { These results can be naturally extended to more general log-likelihood functions.}
\end{remark}

\def\Minv{M^{-1}}
\def\subTconj{_{\widetilde{\Tbb}}}
\def\bmuhat{\widehat{\bmu}}
\def\Tsc{\mathcal{T}}
\def\Esc{\mathcal{E}}
\def\lambdatilde{\tilde{\lambda}}
\def\Cscr{\mathscr{C}}
\def\Ccomp{\Cscr_{\sf\scriptscriptstyle comp}}
\def\Cirr{\Cscr_{\sf\scriptscriptstyle Irrep}}
\def\bgamma{\boldsymbol{\gamma}}
\def\subalpha{_{\balpha}}
\def\submu{_{\bmu}}
\def\Gscr{\mathscr{G}}
\def\Wbbhat{\Wbb}
\def\supt{^{\scriptscriptstyle \sf (t)}}
\def\supMmone{^{\scriptscriptstyle \sf (M-1)}}
\def\supMmonetrans{^{\scriptscriptstyle \sf (M-1) T}}
\def\bI{{\bf I}}
\def\bdiag{{\sf bdiag}}

%%%%%%%%%%%%%%%%%%%%%%%%%%%%%%%%%%%%%%%
%%%%%%%%%%%%%%%%%%%%%%%%%%%%%%%%%%%%%%%
\section{Theoretical Results}\label{sec:thm}
%\tcomm{can you clean up your notation to match the earlier section}
%%%%%%%%%%%%%%%%%%%%%%%%%%%%%%%%%%%%%%%
%%%%%%%%%%%%%%%%%%%%%%%%%%%%%%%%%%%%%%%
In this section, we present theoretical properties of $\bbetahat\subadel\supbullet$ for $\rho(\bbeta\supbullet) = \rho_2(\bbeta\supbullet)$ but discuss how our theoretical results can be extended to other sparse structures in Section \ref{sec:dis}. In Sections \ref{sec:pred} and \ref{sec:equiv}, we derive theoretical consistency and equivalence for the prediction and estimation risks of the SHIR, under high dimensional sparse model and smooth loss function $f$. { In Section \ref{sec:com:deb}, we compare the risk bounds for SHIR with an estimator derived based on those of the debiasing-based aggregation approaches \citep{lee2017communication,battey2018distributed}.}  In addition, Section \ref{sec:spa} shows that the SHIR  achieves sparsistency, i.e., variable selection consistency, for the non-zero sets of $\bmu_0$ and $\balpha\supbullet_0$. We begin with some notation and definitions that will be used throughout the paper.

%%%%%%%%%%%%%%%%%%%%%%%%%%%%%%%%%%%%%%%
\subsection{Notation and definitions}\label{sec:not_thm}
%%%%%%%%%%%%%%%%%%%%%%%%%%%%%%%%%%%%%%%
{ 
Let $o\{\alpha(n)\}$, $O\{\alpha(n)\}$, $\omega\{\alpha(n)\}$, $\Omega\{\alpha(n)\}$ and $\Theta\{\alpha(n)\}$ respectively represent the sequences that grow in a smaller, equal/smaller, larger, equal/larger and equal rate of the sequence $\alpha(n)$. Similarly, we let $o_{\sf P}$, $O_{\sf P}$, $\omega_{\sf P}$, $\Omega_{\sf P}$ and $\Theta_{\sf P}$ represent each of the corresponding rates with probability approaching $1$ as $n\rightarrow \infty$.} We define the {\em model complexity adjusted effective} sample size for each study as {$n_m\supeff = n_m/(s_0\log p)$ and $\neff = N/[s_0(\log p+M)]$}, which are the main drivers for the rates of the proposed estimators. Following \cite{vershynin2018high}, we define the sub-Gaussian norm of a random variable $X$ as $\|X\|_{\psi_2}:=\sup_{q\geq1}q^{-1/2}(\ep|X|^q)^{1/q}$. For any symmetric matrix $\Xbb$, let $\Lambda_{\min}(\Xbb)$ and $\Lambda_{\max}(\Xbb)$ denote its minimum and maximum eigenvalue respectively. Let $\Tbb=(\bone_{(M-1)\times1},\Ibb_{(M-1)\times(M-1)})\trans$ and define $\|\bx\|_{\Tbb}:=\|\Tbb\bx\|_2$ for $\bx\in\mathbb{R}^{M-1}$ and its conjugate norm as $\|\bx\|\subTconj:=\|\Tbb(\Tbb\trans\Tbb)^{-1}\bx\|_2$. For $a\in\mathbb{R}$, denote by ${\rm sign}(a)$ the sign of $a$, and for event $\Esc$, denote by $\bI(\Esc)$ the indicator for $\Esc$. Denote by $\Ssc_{\mu}=\{j:\mu_{0j}\neq0\}$, $\Ssc_{\alpha}=\{j:\|\balpha_{0j}\|_2\neq0\}$, $\Ssc_0=\Ssc_{\mu}\cup\Ssc_{\alpha}$, $s_{\mu}=|\Ssc_{\mu}|$, $s_{\alpha}=|\Ssc_{\alpha}|$ and $s_0=|\Ssc_0|$. Let $f'_1(a,y) = \partial f(a,y)/\partial a$ and $f''_1(a,y) = \partial^2 f(a,y)/\partial a^2$.  

{
The weighted design matrix corresponding to $\Lschat\subadel(\bmu,\balpha\supbullet)$ with respect to $\btheta = (\bmu, \balpha\suptwotrans, \ldots, \balpha\supMtrans)\trans$ after setting $\balpha\supone = - \sum_{m =2}^M \balpha\supm$ can be expressed as
$$
\Wbb(\bbeta\supbullet)= {\sf bdiag}\{\O_1^{1/2}(\bbeta\supone),\ldots,\O_M^{1/2}(\bbeta\supM)\} \Zbb,
$$
where ``${\sf bdiag}$'' is the block diagonal operator, $\O_m(\bbeta)={\sf diag}\{f''_1(\bbeta\trans \bX_1\supm,Y\supm _1),\ldots,f''_1(\bbeta\trans\bX_{n_m}\supm ,Y\supm _{n_m})\}$,  $\Zbb=\Zbb_{[p],[p]}$, and for any $\Ssc_{1}, \Ssc_{2} \subseteq [p]$,
\begin{equation*}
\Zbb_{\Ssc_1,\Ssc_2}=\left(\begin{matrix}
\Xbb_{\bullet\Ssc_1}\supone&-\Xbb_{\bullet\Ssc_2}\supone&-\Xbb_{\bullet\Ssc_2}\supone&\cdots&-\Xbb_{\bullet\Ssc_2}\supone\\
\Xbb_{\bullet\Ssc_1}\suptwo&\Xbb_{\bullet\Ssc_2}\suptwo&\mathbf{0}&\cdots&\mathbf{0}\\
\Xbb_{\bullet\Ssc_1}\supthree&\mathbf{0}&\Xbb_{\bullet\Ssc_2}\supthree&\cdots&\mathbf{0}\\
\vdots&\vdots&\vdots&\ddots&\vdots\\
\Xbb_{\bullet\Ssc_1}\supM &\mathbf{0}&\mathbf{0}&\cdots&\Xbb_{\bullet\Ssc_2}\supM
\end{matrix}\right) .
\end{equation*}
%with $\Xbb_{\bullet\Ssc_j}\supm$ representing the columns of $\Xbb\supm$ corresponding to the set $\Ssc_j$ for $j=1,2$.
%Note that $\Wbbhat(\bbetahat\supbullet\sublasso)$ represents the weighted design matrix for the loss $\Lschat\subadel(\bmu,\balpha\supbullet)$ as a function of $(\bmu\trans,\balpha\suptwotrans,\ldots,\balpha\supMtrans)\trans$ with $\balpha\supone$ substituted by $-\sum_{m\ne 1}\balpha\supm$. 
For any $\Ssc_{1}, \Ssc_{2} \subseteq [p]$, let $\Hbb_{m,\Ssc_1}(\bbeta\supm)$ represent the sub-matrix of $\Hbb_m(\bbeta\supm):=\nabla^2\Lschat_m(\bbeta\supm)$ with its rows and columns corresponding to $\Ssc_1$,   and $\Wbb_{\Ssc_1,\Ssc_2}(\bbeta\supbullet)$ denote the sub-matrix of $\Wbb(\bbeta\supbullet)$ corresponding to $\Zbb_{\Ssc_1,\Ssc_2}$ and $(\bmu\trans_{\Ssc_1},\balpha\suptwotrans_{\Ssc_2},\ldots,\balpha_{\Ssc_2}\supMtrans)\trans$. For notational ease, let $\Hbbhat={\sf bdiag}\{\Hbbhat_1,\Hbbhat_2,\ldots,\Hbbhat_M\}$, $\Ssc_{\sf full}=\{\Ssc_{\mu},\Ssc_{\alpha}\}$ and $\Wbbhat\subSfull(\bbeta\supbullet)=\Wbbhat_{\Ssc_{\mu},\Ssc_{\alpha}}(\bbeta\supbullet)$. Let $\Hbb(\bbeta\supbullet)={\sf bdiag}\{\Hbb_1(\bbeta\supone),\Hbb_2(\bbeta\suptwo),\ldots,\Hbb_M(\bbeta\supM)\}$. We next introduce the compatibility condition and irrepresentable condition below.}

{
\begin{definition}
\emph{\textbf{Compatibility Condition} ($\Ccomp$):} 
The Hessian matrix $\Hbb(\bbeta\supbullet)$ and the  index set $\mathcal{S}$ satisfy the Compatibility Condition with constant $t$, if there exists constant $\phi_0\{t,\mathcal{S},\Hbb(\bbeta\supbullet)\}$ such that for all $(\bmu_{\Delta}\trans ,\balpha\supbullettrans_{\Delta})\trans =(\bmu\trans _{\Delta},\balpha\suponetrans_{\Delta},\ldots,\balpha\supMtrans_{\Delta} )\trans \in \Csc(t,\mathcal{S})$,
\[
(\|\bmu_{\Delta}\|_1+\lambda_g\|\balpha\supbullettrans_{\Delta}\|_{2,1})^2\leq\Ninv \sum_{m=1}^Mn_m(\bmu_{\Delta}+\balpha_{\Delta}\supm )\trans \Hbb_m(\bbeta\supm) (\bmu_{\Delta}+\balpha_{\Delta}\supm )|\mathcal{S}|/\phi_0\{t,\mathcal{S},\Hbb(\bbeta\supbullet)\},
\]
where $\Csc(t,\mathcal{S})=\{(\u\trans ,\v\supbullettrans )\trans= (\u\trans ,\v\suponetrans,\ldots,\v\supMtrans)\trans:\v\supone+\cdots+\v\supM=\mathbf{0},~\|\u_{\mathcal{S}^c}\|_1+\lambda_g\|\v\supbullet_{\mathcal{S}^c}\|_{2,1}\leq t(\|\u_\mathcal{S}\|_1+\lambda_g\|\v\supbullet_{S}\|_{2,1})\}$ for any $t$ and $\Ssc$, and $\phi_0\{t,\mathcal{S},\Hbb(\bbeta\supbullet)\}$ represents the compatibility constant of $\Hbb(\bbeta\supbullet)$ on the set $\mathcal{S}$.
\label{def:1}
\end{definition}

\begin{definition}
\emph{\textbf{Irrepresentable Condition} ($\Cirr$):} The design matrix $\Wbbhat(\bbeta\supbullet)$ satisfies the Irrepresentable Condition on $\Ssc_{\sf full}=(\Ssc_{\mu},\Ssc_{\alpha})$ with parameter $\epsilon>0$, if for all $j\in \mathcal{S}_{\mu}^c$ and $j'\in \mathcal{S}_{\alpha}^c$,
\begin{align*}
&\sup_{\u\in\Gscr_{\Ssc_{\mu}},\v\supbullet\in\Gscr_{\Ssc_{\alpha}}}\left\{\left|(\u\trans,\lambda_g\v\supbullettrans) \left[\Wbbhat\trans_{\mathcal{S}_{\sf full}}(\bbeta\supbullet)\Wbbhat_{\mathcal{S}_{\sf full}}(\bbeta\supbullet)\right]^{-1}\Wbbhat\trans_{\mathcal{S}_{\sf full}}(\bbeta\supbullet)\Wbbhat_{j,\emptyset}(\bbeta\supbullet)\right|\right\}\leq1-\epsilon;\\
& \sup_{\u\in\Gscr_{\Ssc_{\mu}},\v\supbullet\in\Gscr_{\Ssc_{\alpha}}}\left\{\left\|(\u\trans,\lambda_g\v\supbullettrans) \left[\Wbbhat\trans_{\mathcal{S}_{\sf full}}(\bbeta\supbullet)\Wbbhat_{\mathcal{S}_{\sf full}}(\bbeta\supbullet)\right]^{-1}\Wbbhat\trans_{\mathcal{S}_{\sf full}}(\bbeta\supbullet)\Wbbhat_{\emptyset,j'}(\bbeta\supbullet)\right\|_{\widetilde{\Tbb}}\right\}\leq\lambda_g(1-\epsilon),
\end{align*}
where 
\begin{align*}
\Gscr_{\Ssc_{\mu}}=&\left\{\u=(u_1,\cdots,u_{|\Ssc_{\mu}|})\trans\in\mathbb{R}^{|\Ssc_{\mu}|}:\max_{j\in[|\Ssc_{\mu}|]}|u_j|\leq 1\right\},\\
\Gscr_{\Ssc_{\alpha}}=&\left\{\v\supbullet=(\v\suptwotrans,\ldots,\v\supMtrans)\trans\in\mathbb{R}^{(M-1)|\Ssc_{\alpha}|}:\max_{j\in[|\Ssc_{\alpha}|]}\|\v_j\|_{\widetilde{\Tbb}}\leq 1,~\v_j=(v_j\suptwo,\ldots,v_j\supM)\trans\right\}
\end{align*}
%\tcomm{should $\Gscr_{\Ssc_{\alpha}}$ be using $(\v\suptwotrans,\ldots,\v\supMtrans)\trans$ instead?}
represent the sub-gradient space corresponding to $\Ssc_{\mu}$ and $\Ssc_{\alpha}$ of the mixture penalty.
\label{def:2}
\end{definition}
}
%%%%%%%%%%%%%%%%%%%%%%%%%%%%%%%%%%%%%%%
\subsection{Prediction and Estimation Consistency}\label{sec:pred}
%%%%%%%%%%%%%%%%%%%%%%%%%%%%%%%%%%%%%%%
To establish theoretical properties of the SHIR estimators in terms of estimation and prediction risks, we first introduce some sufficient conditions. Throughout the following analysis, we assume that $n_m=\Theta(N/M)$ for $m\in[M]$ and $\lambda_g=\Theta(M^{-1/2})$.
%\begin{cond}
%There exists positive constants $C_{\min}$ and $C_{\max}$ that for any $m\in[M]$, $C_{\min}\leq\Lambda_{\min}(\S_{0,-1}\supm )\leq\Lambda_{\max}(\S_{0,-1}\supm )\leq C_{\max}$ and $C_{\min}\leq\Lambda_{\min}\{\S_0\supm(\bbeta\supm_0)\}\leq\Lambda_{\max}\{\S_0\supm(\bbeta\supm_0)\}\leq C_{\max}$.
%\label{cond:1}
%\end{cond}
{
%{\purple [if change the following condition to a neighborhood version, we may remove the cross-fitting statements in the main text. move the statements in Section \ref{sec:SHIR } to the Supplement A.1.]}
\begin{cond}
There exists {$\delta_1=\omega\{(1/n_m\supeff)^{1/2}\}$} and $\phi_0=\Theta(1)$ such that for all $\bbeta\supbullet=(\bbeta\suponetrans,\ldots,\bbeta\supMtrans)\trans$ satisfying $\|\bbeta\supm-\bbeta_0\supm\|_2<\delta_1$ and $t=\Theta(1)$, the Hessian matrices $\Hbb(\bbeta\supbullet)$ and the index set $\Ssc_0$ satisfy $\Ccomp$ (Definition \ref{def:1}) with compatibility constant $\phi_0\{t,\mathcal{S}_0,\Hbb(\bbeta\supbullet)\}\geq\phi_0$.
\label{cond:1}
\end{cond}

\begin{cond}
For all $m\in[M]$, $X\supm_{ij}f_1'(\bbeta\supmtrans_0\bX\supm_i,\bY\supm_i)$ is sub-Gaussian, i.e. there exists some positive constant $\kappa=\Theta(1)$ such that $\|X\supm_{ij}f_1'(\bbeta\supmtrans_0\bX\supm_i,\bY\supm_i)\|_{\psi_2}<\kappa$. In addition, there exists $B>0$ such that $\max_{m\in[M],i\in[n_m]}\|\bX\supm_i\|_{\infty}\leq B$.
\label{cond:2}
\end{cond}
%{\blue [I modify the condition a bit, I think it was a typo, please double check. Yin]}

\begin{cond}
There exists positive $C_L=\Theta(1)$ such that $|f''_1(a,y)-f''_1(b,y)|\leq C_L|a-b|$ for all $a,b\in\mathbb{R}$.
\label{cond:3}
\end{cond}

\begin{remark}
{Condition \ref{cond:1} holds in the same spirit as the restricted strong convexity (RSC) condition corresponding to the general decomposable penalty introduced in \cite{negahban2012unified}. In Appendix \ref{sec:sup:comp} of the Supplementary Material, we further illustrate that Condition \ref{cond:1} is not stronger than the RSC condition.} The first part of Condition \ref{cond:2} controls the tail behavior of $X\supm_{ij}f_1'(a,y)$ so that the random error $\nabla \Lschat_m(\bbeta_0\supm )$ can be bounded properly and the method could be benefited from the group sparsity of $\balpha\supbullet$ \citep{huang2010benefit}. This condition can be easily verified for sub-gaussian design and an extensive class of models, e.g. the logistic model. In addition, the condition $\max_{m\in[M],i\in[n_m]}\|\bX\supm_i\|_{\infty}\leq B$ holds for bounded design with $B=\Theta(1)$ and for sub-gaussian design with $B=\Theta[\{\log (pN)\}^{1/2}]$. Condition \ref{cond:3} assumes a smooth function $f$ to guarantee that the empirical Hessian matrix $\nabla^2 \Lschat_m(\bbetahat\sublasso\supm)$ is close enough to $\nabla^2 \Lschat_m(\bbeta_0\supm )$, and the term
$\bghat_m=[\Hbbhat_m\bbetahat\sublasso\supm-\nabla \Lschat_m(\bbetahat\sublasso\supm)]$ is close enough to $[\nabla^2\Lschat_m(\bbeta_0\supm)\bbeta_0\supm-\nabla \Lschat_m(\bbeta_0\supm)]$. %\tcomm{we now only consider cross-fitted estimators for $\Hbbhat_m$ and $\bghat_m$. should we use those instead?} 
%{For Condition \ref{cond:3} {(iii)}, $\sqrt{\log p}$ is assumed to be a constant to simplify the technical details but it is not a necessary condition to attain the risk optimality. As an example, if $\sqrt{\log p}$ grows in the order of $(\log p)^{\half}$ for sub-Gaussian design $\Xbb\supm$, we can simply modify the sparsity assumption accordingly to attain the optimality, as $\sqrt{\log p}$ only appears in the excess term beyond the statistical minimax rate as shown in Theorem \ref{thm:1}.} %In practice, $\sqrt{\log p}=O(1)$ is satisfied for categorical design and can be realized by standardizing the columns of $X\supm$ or normalizing the columns of $X\supm$ to be $1$.
\label{rem:1}
\end{remark}

We further assume in Condition \ref{cond:4} that the local LASSO estimators achieve the minimax optimal error rates to a logarithmic scale \citep{raskutti2011minimax}. 
\begin{cond}
The local estimators satisfy that $\max_{m\in[M]}\|\bbetahat\sublasso\supm-\bbeta_0\supm\|_1=O_{\sf P}\{(s_0/n_m\supeff)^{1/2}\}$, $\max_{m\in[M]}\|\bbetahat\sublasso\supm-\bbeta_0\supm\|_2=O_{\sf P}\{(1/n_m\supeff)^{1/2}\}$ and $\max_{m\in[M]}\|\Xbb\supm(\bbetahat\sublasso\supm-\bbeta_0\supm)\|_2=O_{\sf P}\{(n_m/n_m\supeff)^{1/2}\}$.
\label{cond:4}
\end{cond}

\begin{remark}
Extensive literatures, such as \cite{van2008high}, \cite{buhlmann2011statistics} and \cite{negahban2012unified}, have established a complete theoretical framework regarding to this property. See, for example, \cite{negahban2012unified}, in which Condition \ref{cond:4} can be proved under sub-gaussian design $\Xbb\supm$ and strongly convex loss function $f$.
\label{rem:loc}
\end{remark}
}
%Before presenting the risk bounds for the proposed procedure, we begin with the theoretical analysis for the $l_1$ difference between the local estimators $\bbetahat\sublasso\supm$ and the truth $\bbeta_0\supm$, as well as the prediction risk for each local cite.
%\begin{lemma}
%\emph{\textbf{(Risk bounds for Local Estimators)}} Under Conditions \ref{cond:1}-\ref{cond:3} and assume $n_m\supeff=\omega(1)$, there exists positive constants $\lambda_m= \Theta\{(s_0n_m\supeff)^{-1/2}\}$ that 
%for all $m\in[M]$,
%\begin{align*}
%&\max_{m\in[M]}\|\bbetahat\sublasso\supm-\bbeta_0\supm\|_1=O_{\sf P}\{(s_0/n_m\supeff)^\half\};\\
%&\max_{m\in[M]}\|\Hbbhat_m^{\frac{1}{2}}(\bbetahat\sublasso\supm-\bbeta_0\supm)\|_{2}=O_{\sf P}\{(1/n_m\supeff)^\half\}.
%\end{align*}
%\label{lem:1}
%\end{lemma}\vspace{-.3in}

Next, we present the risk bounds for the SHIR including the prediction risk $\|\Hbbhat^{1/2}(\bbetahat\subadel\supbullet-\bbeta_0\supbullet)\|_2$ and estimation risk $\|\widehat{\bmu}\subadel-\bmu_0\|_1+\lambda_g\|\widehat{\balpha}\subadel\supbullet-{\balpha}_{0}\supbullet\|_{2,1}$.
{
\begin{theorem}
\emph{\textbf{(Risk bounds for the SHIR)}} 
Under Conditions \ref{cond:1}--\ref{cond:4} and assume $n_m=\Theta(N/M)$ for all $m\in[M]$. There exists $\lambda=\Theta\{1/(s_0 \neff)^{1/2}+ B/n\supeff_m\}$ and $\lambda_g=\Theta(M^{-1/2})$ such that
\begin{align*}
&\|\Hbbhat^{\frac{1}{2}}(\bbetahat\subadel\supbullet-\bbeta_0\supbullet)\|_2= O_{\sf P}\{(1/\neff)^{\frac{1}{2}}+Bs_0^{\half}/{n_m\supeff}\};\\
&\|\widehat{\bmu}\subadel-\bmu_0\|_1+\lambda_g\|\widehat{\balpha}\subadel\supbullet-{\balpha}_{0}\supbullet\|_{2,1}= O_{\sf P}\{(s_0/\neff)^{\frac{1}{2}}+Bs_0/n_m\supeff\}.
\end{align*}
\label{thm:1}
\end{theorem}\vspace{-.4in}
}
%\begin{remark}
%{\blue
%By Lemma \ref{lem:1}, the condition $\bbetahat\sublasso\supm \in \Bscr_{r}(\bbeta_0\supm)$, i.e. $\|\bbetahat\sublasso\supm-\bbeta_0\supm \|_2<r$ in Theorem \ref{thm:1} holds when 
%$s_0=o\{N/(M\log p)\}$, or equivalently $n_m\supeff\rightarrow\infty$ as $N\rightarrow\infty$, which is a mild sparsity assumption.
%}
%\label{rem:2}
%\end{remark}
%Theorem \ref{thm:1} gives the upper bounds for the prediction risk and the estimation risk of the SHIR  $\bbetahat\subadel\supbullet$. 
\noindent The second term in each of the upper bounds of Theorem \ref{thm:1} is the error incurred by aggregation noise of derived data instead of raw data. These terms are asymptotically negligible under sparsity as $s_0=o(\{N(\log p+M)\}^{1/2}/[BM\log p])$. Then $\bbetahat\subadel\supbullet$ achieves the same error rate as the ideal estimator $\bbetahat\supbullet\subIPD$ obtained by combining raw data as shown in the following section, and is nearly rate optimal. 

%%%%%%%%%%%%%%%%%%%%%%%%%%%%%%%%%%%%%%%
\subsection{Asymptotic Equivalence in Prediction and Estimation}\label{sec:equiv}
%%%%%%%%%%%%%%%%%%%%%%%%%%%%%%%%%%%%%%%
{
Under specific sparsity assumptions, we show the asymptotic equivalence, with respect to prediction and estimation risks, of the SHIR and the ideal IPD estimator $\bbetahat\subIPD\supbullet$ or alternatively defined as
\[
(\widehat{\bmu}\subIPD,\widehat{\balpha}\supbullet\subIPD)=\argmin{(\bmu,\balpha\supbullet)}~\Lschat(\bmu,\balpha\supbullet)+\lambdatilde\rho_2(\bmu,\balpha\supbullet;\lambda_g),~~\mbox{s.t.}~\bone_{M\times 1}\trans \balpha_j=0,~j\in[p],
\]
where $\lambdatilde$ is a tuning parameter.
}
\begin{theorem}
\emph{\textbf{(Asymptotic Equivalence)}} Under assumptions in Theorem \ref{thm:1} and assume $s_0=o(\{N(\log p+M)\}^{1/2}/[BM\log p])$, there exists $\lambdatilde=\Theta\{1/(s_0 \neff)^{1/2}\}$ and $\lambda_g=\Theta(1/M^{1/2})$ such that the IPD estimator $\bbetahat\subIPD\supbullet$ satisfies
\begin{align*}
&\|\Hbbhat^{\frac{1}{2}}(\bbetahat\subIPD\supbullet-\bbeta_0\supbullet)\|_2= O_{\sf P}\{(1/\neff)^{\frac{1}{2}}\};\\
&\|\widehat{\bmu}\subIPD-\bmu_0\|_1+\lambda_g\|\widehat{\balpha}\subIPD\supbullet-{\balpha}_{0}\supbullet\|_{2,1}= O_{\sf P}\{(s_0/\neff)^{\frac{1}{2}}\}.
\end{align*}
Furthermore, for some $\lambda_{\Delta}=o(\lambdatilde)$, the IPD and the SHIR defined by (\ref{equ:loss}) with $\lambda=\lambdatilde+\lambda_{\Delta}$ are equivalent in prediction and estimation in the sense that
\begin{align*}
&\|\Hbbhat^{\frac{1}{2}}(\bbetahat\subadel\supbullet-\bbeta_0\supbullet)\|_2\leq \|\Hbbhat^{\frac{1}{2}}(\bbetahat\supbullet\subIPD-\bbeta_0\supbullet)\|_2+o_{\sf P}\{(1/\neff)^{\frac{1}{2}}\};\\
&\|\widehat{\bmu}\subadel-\bmu_0\|_1+\lambda_g\|\widehat{\balpha}\subadel\supbullet-{\balpha}_{0}\supbullet\|_{2,1}\leq\|\widehat{\bmu}\subIPD-\bmu_0\|_1+\lambda_g\|\widehat{\balpha}\subIPD\supbullet-{\balpha}_{0}\supbullet\|_{2,1}+o_{\sf P}\{(s_0/\neff)^{\frac{1}{2}}\}.
\end{align*}
\label{thm:2}
\end{theorem}\vspace{-.3in}
\noindent Theorem \ref{thm:2} demonstrates the asymptotic equivalence between $\bbetahat\subadel\supbullet$ and $\bbetahat\supbullet\subIPD$ with respect to estimation and prediction risks, and hence implies strict optimality of the SHIR. %In addition, we provide a novel framework for the proof of Theorem \ref{thm:2} given in Section \ref{sec:proof:thm2}.
%{\blue Neither do we find similar proof scheme as our proof for theorem \ref{thm:2} presented in section \ref{sec:proof:thm2} in existing literatures.}
%%
Specifically, when $s_0=o(\{N(\log p+M)\}^{1/2}/[BM\log p])$, the excess risks of $\bbetahat\subadel\supbullet$ compared to $\bbetahat\supbullet\subIPD$ are of smaller order than those of IPD, i.e. the minimax optimal rates (up to a logarithmic scale) for multi-task learning of high dimensional sparse model \citep{lounici2011oracle,huang2010benefit}. %The additional noise of the SHIR  approach arises from the difference between $\bghat_m$ and $n_m[\nabla^2\Lschat_m(\bbeta\supbullet_0)\bbeta\supbullet_0-\nabla \Lschat_m(\bbeta\supbullet_0)]$, which can be controlled by employing a slightly larger tuning $\lambda=\lambda_{\Delta}+\lambda_r>0$. 
Similar equivalence results was given in Theorem 4.8 of \cite{battey2018distributed} for the truncated debiased LASSO estimator. However, to the best of our knowledge, in the existing literatures, such results have not been established yet for the LASSO-type estimators obtained directly from a sparse regression model. Compared with \cite{battey2018distributed}, our result does not require the Hessian matrix $\Hbbhat_m$ to have a sparse inverse since we do not actually rely on the debiasing of $\bbetahat\supm\sublasso$. Consequently, the proofs of Theorem \ref{thm:2} are much more involved than those in \cite{battey2018distributed}. The technical difficulties are briefly discussed in Section \ref{sec:dis} and new technical skills are developed and presented in detail in the Supplementary Material.

\def\supplus{^{\scriptscriptstyle +}}
\def\supsplus{^{\scriptscriptstyle s_+}}
\def\suphplus{^{\scriptscriptstyle h_+}}
\def\sign{\mbox{sign}}
{ 
%%%%%%%%%%%%%%%%%%%%%%%%%%%%%%%%%%%%%%%
\subsection{Comparison with the debiasing-based strategy}\label{sec:com:deb}
%%%%%%%%%%%%%%%%%%%%%%%%%%%%%%%%%%%%%%%
To compare to existing approaches, we next consider an extension of the debiased LASSO based procedures proposed in \cite{lee2017communication} and \cite{battey2018distributed} to incorporating between study heterogeneity.  Specifically, at the $m\supth$ site, we derive the debiased LASSO estimator $\bbetahat\supm\subdlasso$ as defined in (\ref{equ:debias}) and send it to the central site, where $\bThetahat_m$ is obtained via nodewise LASSO \citep{javanmard2014confidence}. At the central site, compute $\bmuhat\subdlasso=\Minv\sum_{m=1}^M\bbetahat\supm\subdlasso$, $\balphahat\supm\subdlasso=\bbetahat\supm\subdlasso-\bmuhat\subdlasso$ and $\balphahat\supbullet\subdlasso=(\balphahat\supone\subdlasso,\ldots,\balphahat\supM\subdlasso)\trans$. The final estimator for $\bmu$ and $\balpha$ can be obtained by thresholding $\bmuhat\subdlasso$ and $\balphahat\supbullet\subdlasso$ as
$\bmuhat\sublab=\Tscr_{\mu}(\bmuhat\subdlasso; \tau_1)$ and $\balphahat\sublab\supbullet=\Tscr_{\alpha}(\balphahat\subdlasso\supbullet; \mu_2)$, by \cite{lee2017communication} and \cite{battey2018distributed}, where
\begin{alignat*}{2}
\Tscr_{\mu}(\bmu; \tau_1)&=\{\mu_1,\mu_2\suphplus(\tau_1), \ldots, \mu_p\suphplus(\tau_1)\}\trans & \quad \mbox{or}\quad  &\{\mu_1,\mu_2\supsplus(\tau_1),\ldots, \mu_p\supsplus(\tau_1)\}\trans \\
\Tscr_{\alpha}(\balpha\supbullet; \tau_2) & = \mbox{vec}\{[\balpha_1,\balpha_2\suphplus(\tau_2),\ldots\balpha_p\suphplus(\tau_2)]\trans\} & \quad \mbox{or} \quad & 
\mbox{vec}\{[\balpha_1,\balpha_2\supsplus(\tau_2),\ldots\balpha_p\supsplus(\tau_2)]\trans\},
\end{alignat*}
for any vector $\bx=(x_1, ..., x_d)\trans$ and constant $\tau$,  $\bx\suphplus = \bx I(\|\bx\|_2 > \tau)$ and $\bx\supsplus = \bx (1- \|\bx\|_2^{-1}\tau)I(\|\bx\|_2 > \tau)$ respectively denote the hard and soft thresholded counterparts of $\bx$, and  $\mbox{vec}(\Abb)$ vectorize the matrix $\Abb$ by column. %{\blue Here, both the hard and soft thresholding versions are considered by \cite{lee2017communication} while only the hard one is considered in \cite{battey2018distributed}. These two strategies have the same error rate as shown in \cite{lee2017communication}.}

The error rates of $\{\bmuhat\sublab,\balphahat\sublab\supbullet\}$ or their corresponding $\bbetahat\sublab\supbullet$ can be largely derived by extending the Theorem 21 results of \cite{lee2017communication}, and Theorem 4.8 of \cite{battey2018distributed}. We outline the results below and provide additional details in Section \ref{sec:sup:debias} of the Supplementary Material. Denote by $\bar\Hbb_{m}(\bbeta\supm)=\ep [\Hbb_m(\bbeta\supm)]$, $\bar\Hbb_{m}=\bar\Hbb_{m}(\bbeta_0\supm)$, $\bar\bTheta_{m}=\{\bar\theta_{mj\ell}\}_{p\times p}=\bar\Hbb_{m}^{-1}$ %\tcomm{I removed "0" from subscript for notational ease} 
and its row-wise sparsity level $s_1=\max_{m\in [M]j\in [p]}|\{\ell\neq j:\bar\theta_{mj\ell}\neq 0\}|$. Then in analog to Theorem \ref{thm:1}, under the same regularity conditions as SHIR, one can obtain that
\begin{equation}
%\begin{split}
%&\|\Hbbhat^{\frac{1}{2}}(\bbetahat\sublab\supbullet-\bbeta_0\supbullet)\|_2= O_{\sf P}\{(1/\neff)^{\frac{1}{2}}+B(s_0^{\half}+s_1/s_0^{\half})/{n_m\supeff}\};\\
\|\widehat{\bmu}\sublab-\bmu_0\|_1+\lambda_g\|\widehat{\balpha}\sublab\supbullet-{\balpha}_{0}\supbullet\|_{2,1}= O_{\sf P}\{(s_0/\neff)^{\frac{1}{2}}+B(s_0+s_1)/n_m\supeff\},
%\end{split}    
\label{equ:deb:error}
\end{equation}
where $B$ is as defined in Condition \ref{cond:2}. Compared with the error rates of SHIR as presented in Theorem \ref{thm:1}, the debiased LASSO based estimator share the same ``first term", i.e. $(s_0/\neff)^{\frac{1}{2}}$, which represents the error of individual level empirical process. However, its second term incurred by data aggregation can be larger than that of SHIR as $s_1=\omega(s_0)$. In many practical settings, $s_1$ could be excessively large due to the complex intrinsic relationship among the covariates, which leads to the inflation of the error rate of $\bmuhat\sublab$ and $\balphahat\sublab\supbullet$. This phenomenon is further validated in our simulation studies and the real example in the following sections.

\begin{remark}
When our model for $Y$ is correctly specified, i.e. ${\sf E}\{f_1'(\bbeta\supmtrans_0\bX\supm_i,\bY\supm_i)\mid\bX\supm_i\}=0$, the risk bound that depends on the exact sparsity level $s_1$ can be potentially relaxed to the one that depends on the approximate sparsity level of $\bar\bTheta_{m}$ \citep{ma2020global,liu2020integrative}, i.e. its row-wise $\ell_q$-norm with $0<q\leq1$. Such sparsity condition seems more reasonable but relies on the model correctness for $Y$, which is another strong assumption that cannot be easily verified in practice. In fact, our following simulation studies demonstrate that the estimators $\bmuhat\sublab$ and $\balphahat\sublab\supbullet$ no longer perform well under the settings with both non-sparse $\bar\bTheta_{m}$ and misspecified models.
\label{rem:debias}
\end{remark}

In addition, SHIR could be more efficient than the debiasing-based strategy even when the impact of the additional error term, which depends on $s_1$ in (\ref{equ:deb:error}), is asymptotically negligible. %This applies when distribution of $\Xbb\supm$'s are heterogeneous across the local sites. 
Consider the setting when all $\bbeta\supm$'s are the same, i.e., $\bbeta\supm=\bbeta$, and $p$ is moderate or small so that the regularization is unnecessary and the maximum likelihood estimator (MLE) for $\bbeta$ is feasible and asymptotically Gaussian. In this case, SHIR can be viewed as the inverse variance weight estimation with asymptotic variance $\boldsymbol{\Sigma}\subadel=\{\sum_{m=1}^Mn_m\bar\bTheta_{m}^{-1}\}^{-1}$, while the debiasing-based approach outputs an estimator of variance $\boldsymbol{\Sigma}\sublab=M^{-2}\sum_{m=1}^Mn_m^{-1}\bar\bTheta_{m}$. It is not hard to show that $\boldsymbol{\Sigma}\subadel\preceq\boldsymbol{\Sigma}\sublab$, where the equality holds only if all $\bar\bTheta_{m}$'s are in certain proportion. Thus, SHIR is strictly more efficient than debiasing-based approach under the low dimensional setting with heterogeneous $\bar\bTheta_{m}$, which commonly arises in meta-analysis as the distributions of $\Xbb\supm$'s are typically heterogeneous across the local sites. %While under the high dimensional setting, the regularized estimators are no longer asymptotic gaussian and such analysis on relative efficiency could not be conducted strictly, due to the lack of tool-kits. However, we still expect 
In the high-dimensional setting, similarly, SHIR is expected to benefit from the ``inverse variance weight" construction, and our simulation results in Section \ref{sec:simu} support this point.

}

%%%%%%%%%%%%%%%%%%%%%%%%%%%%%%%%%%%%%%%
\subsection{Sparsistency}\label{sec:spa}
%%%%%%%%%%%%%%%%%%%%%%%%%%%%%%%%%%%%%%%
In this section, we present theoretical results concerning the variable selection consistency of the SHIR . We begin with some extra sufficient conditions for the sparsistency result.
%In the following analysis, {\blue we still assume that $n_m=\Theta(N/M)$ for $m\in[M]$ and $\lambda_g=\Theta(M^{-1/2})$. } \tcomm{no need to repeat if this is assumed throughout}

{
\begin{cond}
There exists {$\delta_2=\omega\{(1/n_m\supeff)^{1/2}\}$} and $C_{\min}=\Theta(1)$ such that for all $\bbeta\supm$ satisfying $\|\bbeta\supm-\bbeta_0\supm\|_2<\delta_2$, $\Lambda_{\min}\{\Hbb_{m,\Ssc_0}(\bbeta\supm)\}>C_{\min}$.
\label{cond:5}
\end{cond}

\begin{cond}
There exists {$\delta_3=\omega\{(1/n_m\supeff)^{1/2}\}$} and $\epsilon=\Theta(1)$ such that for all $\bbeta\supbullet=(\bbeta\suponetrans,\ldots,\bbeta\supMtrans)\trans$ satisfying $\|\bbeta\supm-\bbeta_0\supm\|_2<\delta_3$, the weighted design matrix $\Wbbhat(\bbeta\supbullet)$ satisfies the Irrepresentable Condition $\Cirr$ (Definition \ref{def:2}) on $\Ssc_{\sf full}$ with constant $\epsilon$.
\label{cond:7}
\end{cond}
}
%{\blue [$\Ssc_{\sf full}$ is not defined anywhere else? Yin]}

\begin{cond}
Let $\nu=\min\{\min_{j\in \mathcal{S}_{\mu}}|\mu_{0j}|,\min_{j\in \mathcal{S}_{\alpha}}\|\balpha_{0j}\|_{2}/M^{1/2}\}$. For the $\epsilon$ defined in Condition \ref{cond:7}, $[1/(\neff)^{1/2}+Bs_0^{1/2}/n_m\supeff]/(\nu\epsilon)\rightarrow 0$, as $N\rightarrow\infty$.
\label{cond:6}
\end{cond}
\begin{remark}
{
Conditions \ref{cond:5}--\ref{cond:6} are sparsistency assumptions similar to those of \cite{zhao2006model} and \cite{nardi2008asymptotic}. Condition \ref{cond:5} requires the eigenvalues {for the covariance matrix of the weighted design matrix} corresponding to $\Ssc_0$ to be bounded away from zero, so that its inverse behaves well. %{\blue Also, it requires $\Hbbhat^{1/2}_{m,\Ssc_0}\bx$ to have at least the same scale as $\Xbb_{\Ssc_0}$, a reasonable regularity condition for models like logistic model and used to remove certain smoothing errors.} 
Condition \ref{cond:7} adopts the commonly used irrepresentable condition \citep{zhao2006model} to our mixture penalty setting. Roughly speaking, it requires that the weighted design corresponding to $\Ssc_{\sf full}$ cannot be represented well by the weighted design for $\Ssc_{\sf full}^c$. Compared to \cite{nardi2008asymptotic}, $\Cirr$ is less intuitive but essentially weaker. We justify such condition on several common hessian structures and compare it with \cite{zhao2006model} in Section \ref{sec:just:irr} of the Supplementary Material. Condition \ref{cond:6} assumes that the minimum magnitude of the coefficients is large enough to make the non-zero coefficients recognizable. As will be discussed later in this section, Condition \ref{cond:6} requires essentially weaker assumption on the minimum magnitude than existing results based on local LASSO \citep{zhao2006model}. This is because we leverage the group structure of $\bbeta\supm_0$'s to improve the efficiency of variable selection.}
\end{remark}

%{\blue In addition, we assume that $\|\Xbb(\bbetahat\subadel\supbullet-\bbeta_0\supbullet)\|_2=O_{\sf P}\{\|\Hbbhat^{\frac{1}{2}}(\bbetahat\subadel\supbullet-\bbeta_0\supbullet)\|_2\}$, a reasonable condition useful in removing some smoothing error.}

\begin{theorem}
\emph{\textbf{(Sparsistency)}} Let $\Sschat_{\mu}=\{j:\muhat_{{\scriptscriptstyle \sf SHIR , }j}\neq0\}$ and $\Sschat_{\alpha}=\{j:\|\widehat{\balpha}_{{\scriptscriptstyle \sf SHIR , }j}\|_2\neq0\}$. Denote the event $\Oscr_{\mu}=\{\Sschat_{\mu}=\mathcal{S}_{\mu}\}$ and $\Oscr_{\alpha}=\{\Sschat_{\alpha}=\mathcal{S}_{\alpha}\}$. Under Conditions \ref{cond:1}--\ref{cond:6} and assume that
\[
\lambda = o(\nu/s_0^{1/2})\quad\mbox{and}\quad\lambda=\epsilon^{-1}\omega(1/(s_0 n\supeff)^{1/2}+ B /n\supeff_m)
\]
we have ${\sf P}(\Oscr_{\mu}\cap\Oscr_{\alpha})\rightarrow1$ as $N\rightarrow\infty$.
\label{thm:3}
\end{theorem}
\begin{remark}
{ Condition \ref{cond:6} guarantees the existence of the tuning parameter $\lambda$ used in Theorem \ref{thm:3}. The range of the tuning parameter $\lambda$ in Theorem \ref{thm:3} is similar to those in \cite{zhao2006model} and \cite{nardi2008asymptotic} in the sense that, $\lambda$ needs to be smaller than the minimum magnitude of the signal divided by $s_0^{1/2}$ to ensure the true positives are selected and dominate the random noise. In comparison to \cite{zhao2006model}, our $\lambda$ additionally includes the consideration of the aggregation noise represented by $B/n\supeff_m$.
}

\end{remark}

{ Theorem \ref{thm:3} establishs the sparsistency of the SHIR estimator. When $s_0=o(\{N(\log p+M)\}^{1/2}/[BM\log p])$ as assumed in Theorem \ref{thm:2}, Condition \ref{cond:6} turns out to be $\nu\epsilon=\omega\{1/(\neff)^{1/2}\}$, the corresponding sparsistency assumption for the IPD estimator. In contrast, a similar condition, which could be as strong as $\nu\epsilon=\omega\{1/(n_m\supeff)^{1/2}\}$, is required for the local LASSO estimator \citep{zhao2006model}. Compared with the local one, our integrative analysis procedure can recognize smaller signal under some sparsity assumptions. In this sense, the structure of $\bbeta\supbullet_0$ helps us to improve the selection efficiency over the local LASSO estimator. Different from the existing work, we need carefully address the mixture penalty $\rho_2$ and the aggregation noise of the SHIR, which introduce technical difficulties to our theoretical analysis.}

In both Theorems \ref{thm:2} and \ref{thm:3}, we allow $M$, the number of studies, to diverge while still preserving theoretical properties. The growing rate of $M$ is allowed to be
\[
M=\min\left(o\{(N/\log p)^{{1}/{2}}/(Bs_0)\},o\{N/(Bs_0\log p)^2\}\right)
\]
for the equivalence result in Theorem \ref{thm:2} and 
{
\[
M=\min\left(o\{N\epsilon\nu/(Bs_0^{{3}/{2}}\log p)\},o\{N(\epsilon\nu)^2/s_0\}\right)
\]
}
for the sparsistency result in Theorem \ref{thm:3}.

%\tcomm{if you let $M$ diverge, we should show that SHIR  does better than single study estimator when beta's are the same}
%\tcomm{Discussion on this has been included after theorem 2, in the blue words.}

\def\expit{\mbox{expit}}
\def\Cbb{\mathbb{C}}
\def\Rbb{\mathbb{R}}
\def\bGamma{\boldsymbol{\Gamma}}

\setlength{\leftmargini}{.0in}

%%%%%%%%%%%%%%%%%%%%%%%%%%%%%%%%%%%%%%%
%%%%%%%%%%%%%%%%%%%%%%%%%%%%%%%%%%%%%%%
\section{Simulation Study}\label{sec:simu}
%%%%%%%%%%%%%%%%%%%%%%%%%%%%%%%%%%%%%%%
%%%%%%%%%%%%%%%%%%%%%%%%%%%%%%%%%%%%%%%
{
We present simulation results in this section to evaluate the performance of our proposed SHIR estimator and compare it with several other approaches. We let $M \in \{4, 8\}$ and $p \in \{100, 800, 1500\}$ and set $n_m=n=400$ for each $m$. For each configuration, we summarize results based on 200 simulated datasets. We consider three data generating mechanisms:

%{\blue [I modified the simulation setting (i) according to the discussion with Tianxi. Molei]}

\begin{itemize}
\item[] (i) {\bf Sparse precision and correctly specified model (strong signal):} Across all studies, we let $\Ssc_{\mu}=\{1,2,\ldots,6\}$ for $\bmu$, $\Ssc_{\alpha}=\{3,4,\ldots,8\}$ for $\balpha$, $\Ssc=\Ssc_{\mu}\cup\Ssc_{\alpha}$ and $\Ssc^c = [p]\setminus \Ssc$. For each $m\in [M]$, we generate $\bX\supm$ from a zero-mean multivariate normal with covariance $\Cbb\supm$, where $\Cbb\supm_{\Ssc^c\Ssc^c} = \Rbb_{p-8}(r_m)$, $\Cbb\supm_{\Ssc^c \Ssc} = \Rbb_{p-8}(r_m)\bGamma_{p-8,8}(r_m,15)$ and $\Cbb\supm_{\Ssc \Ssc}=\mathbb{I}_8+\bGamma\trans_{p-8,8}(r_m,15)\Rbb_{p-8}(r_m)\bGamma_{p-8,8}(r_m,15)$ where $\mathbb{I}_q$ denotes the $q\times q$ identity matrix, $\Rbb_q(r)$ denotes the $q\times q$ correlation matrix of ${\sf AR}(1)$ with correlation coefficient $r$, $\bGamma_{q_1,q_2}(r,s_1)$ denotes the $q_1\times q_2$ matrix with each of its column having randomly picked $s_1$ entries set as $r$ or $-r$ in random and the remaining being $0$, and $r_m=0.4 (m-1)/ M + 0.15$. Given $\bX\supm$, we generate $Y\supm$ from the logistic model $P(Y\supm = 1 \mid \bX\supm) = \expit\{\bX_{\Ssc_{\mu}}\supmtrans\bmu_{\Ssc_{\mu}}+\bX_{\Ssc_{\alpha}}\supmtrans\balpha_{\Ssc_{\alpha}}\supm\}$ with $\bmu_{\Ssc_\mu}=0.5(1,-1,1,-1,1,-1)\trans$ and $\balpha_{\Ssc_\alpha}\supm=0.35(-1)^m\cdot(1, 1,1, -1, -1,-1)\trans$.

\item[] (ii) {\bf Sparse precision and correctly specified model (weak signal):} Use the same data generation mechanism as in (i) except relatively weak signals $\bmu_{\Ssc_\mu}=0.2(1,-1,1,-1,1,-1)\trans$ and $\balpha_{\Ssc_\alpha}\supm=0.15(-1)^m\cdot(1, 1,1, -1, -1,-1)\trans$.

\item[] (iii) {\bf Dense precision and wrongly specified model:} Let $\Ssc=\{1,2,\ldots,5\}$, $\Ssc'=\{6,\ldots,50\}$,  and $\Ssc'' = [p]\setminus (\Ssc\cup\Ssc')$. For each $m \in [M]$, we generate $\bX\supm$ from zero-mean multivariate normal with covariance matrix $\Cbb\supm$, where $\Cbb\supm_{(\Ssc'\cup\Ssc'')(\Ssc'\cup \Ssc'')} = \bdiag\{\Rbb_{45}(r_m), \Rbb_{p-50}(r_m)\}$, $\Cbb\supm_{\Ssc \Ssc''} = \bzero$, $\Cbb\supm_{\Ssc'\Ssc} = \Rbb_{45}(r_m)\bGamma_{45,5}(r_m,45)$ and $\Cbb\supm_{\Ssc\Ssc} =\mathbb{I}_5+\bGamma\trans_{45,5}(r_m,45)\Rbb_{45}(r_m)\bGamma_{45,5}(r_m,45)$. Given $\bX\supm$, we generate $Y\supm$ from a logistic model with $P(Y\supm = 1 \mid \bX\supm) = \expit\{\sum_{j=1}^5\{0.25+0.15(-1)^m\}\{X_j\supm+0.2(X_j\supm)^3\}+0.1\sum_{j=1}^4X_j\supm X_{j+1}\supm\}$.
%{\blue [The simulation settings seem to be very carefully specified, especially the last non-linear link function. I feel the reviewers might ask why we choose this very special setup...may comment on how general it is? Also, we may consider add the comparison with the ``true'' coefficient from the ``true'' model (see AE's Comment 3)? Yin]}
\end{itemize}

Across all settings, the distribution of $\bX\supm$ and model parameters of $Y\supm \mid \bX\supm$ differ across the $M$ sites to mimic the heterogeneity of the covariates and models. The heterogeneity of $\bX\supm$ is driven by the study-specific correlation coefficient $r_m$ in its covariance matrix $\Cbb\supm$. {Under Settings (i) and (ii), the fitted logistic loss corresponds to the likelihood under a correctly specified model with the support of $\bmu$ and that of $\balpha\supm$ overlapping but not exactly the same. Under Setting (iii), the fitted loss corresponds to a mis-specified model but the true target parameter $\bbeta\supm$ remains approximately sparse with only first 5 elements being relatively large, 45 close to zero and remaining exactly zero.} For each $j\in\Ssc$, there are 15 non-zero coefficients on average in the $j$-th column (except $j$ itself) of the precision $\bTheta_{m}$ under Settings (i) and (ii), and 45 non-zero coefficients under Setting (iii). So we can use Settings (i) and (ii) to simulate the scenario with sparse precision on the active set and use Setting (iii) to simulate relatively dense precision. %In addition, the vector $\{\|\balpha_{0j}\|_2, j = 1, ..., p\}$ is approximately sparse with heterogeneity present only for a small subset of covariates.
%the true $\bbeta\supm$'s have their first $50$ elements being non-zero while only the first $5$ coefficients are significant. Thus, one can regard the $\bbeta\supm$ as approximately sparse under this setting, which also enables oracle estimation with LASSO-type regularized regression under high dimensional setting \citep{negahban2012unified}. And the $\bbeta\supm$'s can be viewed as group sparse with heterogeneity due to our construction of the true model. This is actually more close to our real example in that the logistic models of $Y\supm\sim\bX\supm$ are misspecified while the misspecified model is still (approximately) group sparse and heterogeneous across the local sites. 

For each simulated dataset, we obtain the SHIR estimator as well as the following alternative estimators: (a) the IPD estimator $\widehat{\bbeta}\supbullet\subIPD=\argmin{\bbeta\supbullet}\Qschat(\bbeta\supbullet)$; (b) the SMA estimator \citep{he2016sparse}, following the sure independent screening procedure \citep{fan2008sure} that reduces the dimension to $n/(3\log n)$ as recommended by \cite{he2016sparse}; and (c) the debiasing-based estimator $\bbetahat\supbullet\sublab$ as introduced in Section \ref{sec:com:deb}, denoted by Debias$\sublab$. For $\bbetahat\supbullet\sublab$, we used the soft thresholding to be consistent with the penalty used by IPD, SMA and SHIR. We used the BIC to choose the tuning parameters for all methods. 

%since we did not observe significant differences between the hard thresholding and soft thresholding. 

In Figures \ref{fig:sim1a} and \ref{fig:sim1b}, we present the relative average absolute estimation error (rAEE), $\|\bbeta\supbullet-\bbeta_0\supbullet\|_1$, and the relative prediction error (rPE), $\|\Xbb(\bbeta\supbullet-\bbeta_0\supbullet)\|_2$, for each estimator compared to the IPD estimator, respectively.  %\tcomm{relative efficiency is usually defined at the MSE scale, not RMSE scale, just call it relative AEE and relative PE} 
Consistent with the theoretical equivalence results, the SHIR estimator attains very close estimation and prediction accuracy as those of the idealized IPD estimator, with rPE and rAEE around $1.03$ under Setting (i), $1.02$ under Setting (ii), and $1.07$ under Setting (iii). The SHIR estimator is substantially more efficient than the SMA under all the settings, with about $50\%$ reduction in both AEE and PE on average. This can be attributed to the improved performance of the local LASSO estimator $\bbetahat\sublasso\supm$ over the MLE $\breve{\bbeta}\supm$ on sparse models. The superior performance is more pronounced for large $p$ such as $800$ and $1500$, because the screening procedure does not work well in choosing the active set, especially in the presence of correlations among the covariates. Compared with Debias$\sublab$, SHIR also demonstrates its gain in efficiency. Specifically, relative to SHIR, Debias$\sublab$ has $20\%\sim 29\%$ higher AEE and $27\%\sim 42\%$ higher PE under these three settings. 
This is consistent with our theoretical results presented in Section \ref{sec:com:deb} that SHIR has smaller error compared to Debias$\sublab$ due to the heterogeneous hessians and aggregation errors. In addition, compared to Setting (i), the excessive error of Debias$\sublab$ is larger in Setting (iii) where the the inverse Hessian $\bar\bTheta_m$ is relatively dense and the model of $Y$ is misspecified. This is consistent with our conclusion in Section \ref{sec:com:deb} and Remark \ref{rem:debias}.

 {
 In Figure \ref{fig:sim2}, we summarize the average true positive rate (TPR) and false discovery rate (FDR) for recovering the support of $\bbeta\supbullet$ under Settings (i) and (ii) where the logistic model is correctly specified.  Again, SMA performed poorly under both settings with either low TPR or high FDR. Both IPD and SHIR have good support recovery performance with all TPRs above $0.91$ and FDRs below $0.13$ under the strong signal setting, and all TPRs above $0.74$ and FDRs below $0.05$ under the weak signal setting. The IPD and SHIR attained similar TPRs and FDRs with absolute differences less than 0.02 across all settings. Compared with IPD and SHIR, Debias$\sublab$ shows worse performance under both settings. Under Setting (i), the TPR of Debias$\sublab$ is consistently lower than that of SHIR by about 0.13 while the FDR of Debias$\sublab$ is generally higher than that of SHIR except that when $p=100$ Debias$\sublab$ attained very low FDR due to over shrinkage. Under  the weak signal Setting (ii) with $M=4$, Debias$\sublab$ is substantially less powerful than SHIR in recovering true signals with TPR lower by as much as $0.52$ while its average FDR is comparable to that of SHIR. When $M=8$,   Debias$\sublab$ attained comparable TPR as that of SHIR but generally has substantially higher FDR.
}
}
\section{Application to EHR Phenotyping in Multiple Disease Cohorts}\label{sec:real}
%%%%%%%%%%%%%%%%%%%%%%%%%%%%%%%%%%%%%%%
%%%%%%%%%%%%%%%%%%%%%%%%%%%%%%%%%%%%%%%

Linking EHR data with biorepositories containing ``-omics" information has expanded the opportunities for biomedical research \citep{kho2011electronic}.
With the growing availability of these high--dimensional data, the bottleneck in clinical research has shifted from a paucity of biologic data to a paucity of high--quality phenotypic data.
Accurately and efficiently annotating patients with disease characteristics among millions of individuals is a critical step in fulfilling the promise of using EHR data for precision medicine.
Novel machine learning based phenotyping methods leveraging a large number of predictive features have improved the accuracy and efficiency of existing phenotyping methods 
\citep{liao2015methods,yu2015toward}.  

While the portability of phenotyping algorithms across multiple patient cohorts  is of great interest, existing phenotyping algorithms are often developed and evaluated for a specific patient population. To investigate the portability issue and develop EHR phenotyping algorithms for coronary artery disease (CAD) useful for multiple cohorts, \cite{liao2015methods} developed a CAD algorithm using a cohort of rheumatoid arthritis (RA) patients and applied the algorithm to other
disease cohorts using EHR data from Partner's Healthcare System. Here,  { we performed integrative analysis of multiple EHR disease cohorts to jointly develop algorithms for classifying CAD status for four disease cohorts including type 2 diabetes mellitus (DM), inflammatory bowel disease (IBD), multiple sclerosis (MS) 
and RA.} Under the DataSHIELD constraint, our proposed SHIR algorithm enables us to let the data determine if a single CAD phenotyping algorithm can perform well across four disease 
cohorts or disease specific algorithms are needed.  

{ For algorithm training, clinical investigators have manually curated gold standard labels on the CAD status used as the response $Y$}, for $n_1=172$ DM patients, $n_2=230$ IBD patients, $n_3=105$ MS patients and $n_4=760$ RA patients. There are a total of $p = 533$ candidate features including both codified features, narrative features extracted via natural language processing (NLP) \citep{zeng2006extracting}, as well as their two-way interactions. Examples of codified features include demographic information, lab results, medication prescriptions, counts of International Classification of Diseases (ICD) codes and Current Procedural Terminology (CPT) codes. Since patients may not have certain lab measurements and missingness is highly informative, we also create missing indicators for the lab measurements as additional features. Examples of NLP terms include mentions of CAD, current smoking (CSMO), non smoking (NSMO) and CAD related procedures. Since the count variables such as the total number of CAD ICD codes are zero-inflated and skewed, we take $\log(x+1)$ transformation and include ${\bf I}(x>0)$ as additional features for each count variable $x$.

For each cohort, we randomly select 50\% of the observations to form the training set for developing the CAD algorithms and use the remaining 50\% for validation. We trained CAD algorithms based on SHIR, Debias$\sublab$ and SMA. Since the true model parameters are unknown, we evaluate the performance of different methods based on the prediction performance of the trained algorithms on the validation set. We consider several standard accuracy measures including the area under the receiver operating characteristic curve (AUC), the { brier score defined as the mean squared residuals on the validation data} , as well as the  $F$-score at threshold value chosen to attain a false positive rate of 5\% ($F_{5\%}$) and 10\% ($F_{10\%}$), where the $F$-score is defined as the harmonic mean of the sensitivity and positive predictive value. The standard errors of the estimated prediction performance measures are obtained by bootstrapping the validation data. We only report results based on tuning parameters selected with BIC as in the simulation studies but note that the results obtained from AIC are largely similar in terms of prediction performance. Furthermore, to verify the improvement of the performance by combining the four datasets, we include the LASSO estimator for each local dataset (Local) as a comparison.

In Table \ref{tab:var}, we present the estimated coefficients for variables that received non-zero coefficients by at least one of the included methods. 
Interestingly, all integrative analysis methods set all heterogeneous coefficients to zero, suggesting that a single CAD algorithm can be used across all cohorts although 
different intercepts were used for different disease cohorts. The magnitude of the coefficients from SHIR largely agree with the published algorithm with most important
features being NLP mentions and ICD codes for CAD as well as total number of ICD codes which serves as a measure of healthcare utilization. The SMA set all 
variables to zero except for age, non-smoker and the NLP mentions and ICD codes for CAD, while Debias$\sublab$ has more similar support to SHIR. 

%{\blue [interestingly, it is like what would happen to single index model: Quad has good AUC but poor Brier Score. Molei]}

%\tcomm{intercept must be different??we selected no feature that is of form $I(x=0)$?}
%\lcomm{Intercept are different. But our model does not select any $I(x=0)$.}

The point estimates along with their 95\% bootstrap confidence intervals of the accuracy measures  are presented in Figure \ref{fig:1}. The results suggest that SHIR has the best performance across all methods, nearly on all datasets and across all measures. { Among the integrative methods, SMA and Debias$\sublab$ performed much worse than SHIR on all accuracy measures. For example, the AUC with its 95\% confidence interval of the CAD algorithm for the RA cohorts trained via SHIR, SMA and Debias$\sublab$ is respectively 0.93 (0.90,0.95), 0.88 (0.84,0.92) and 0.86 (0.82,0.90).} Compared to the local estimator, SHIR also performs substantially better. For example, the AUC of SHIR and Local for the IBD cohort is 0.93 (0.88,0.97) and 0.90 (0.84,0.95). The difference between the integrative procedures and the local estimator is more pronounced for the DM cohort with AUC being around 0.95 for SHIR and 0.90 for the local estimator trained using DM data only. The local estimator fails to produce an informative algorithm for the MS cohort due to the small size of the training set. These results again demonstrate the power of borrowing information across studies via integrative analysis. 
 
%%%%%%%%%%%%%%%%%%%%%%%%%%%%%%%%%%%%%%%
%%%%%%%%%%%%%%%%%%%%%%%%%%%%%%%%%%%%%%%
\section{Discussion}\label{sec:dis}
%%%%%%%%%%%%%%%%%%%%%%%%%%%%%%%%%%%%%%%
%%%%%%%%%%%%%%%%%%%%%%%%%%%%%%%%%%%%%%%
In this paper, we proposed a novel approach, the SHIR, for integrative analysis of high dimensional data under the DataSHIELD framework, where only summary data is allowed to be transferred from the local sites to the central site to protect the individual-level data. One should notice that the DataSHIELD framework is essentially different from the differential privacy concept \citep{dwork2011differential}. Differential privacy establishes rigorous probabilistic definition on the privacy of a random data processing algorithm, requiring the perturbation of the raw data with random noise \citep{wasserman2010statistical}. It allows for sharing and publishing individual-level data for more general data analysis purposes. However, to ensure enough privacy, the differential privacy was shown to be too strict to be used for high-dimensional regression and inference \citep{duchi2013local}. In comparison, the DataSHIELD framework is used for regression analysis and shown to work well under ultra-high dimensionality, with no efficiency loss compared with the ideal case where the data can be operated with no privacy constraint.

{ 
Though motivated by the debiased LASSO, the SHIR approach does not need to obtain $\bThetahat_m$ or the debiased estimator $\bbetahat\supm\subdlasso$. As we demonstrated via both theoretical analyses and numerical studies, the SHIR estimator is considerably more efficient than the estimators obtained based on the debiasing-based strategies considered in literatures \citep{lee2017communication,battey2018distributed}. The superior performance is even more pronounced when $\bTheta_m$ is relatively dense.} Also, our method accommodates heterogeneity among the design matrices, as well as the coefficients of the local sites, which is not adequately handled under the ultra high dimensional regime in existing literature. In terms of theoretical analysis, no existing distributed learning work provides similar estimation equivalence results as shown in Theorem \ref{thm:2} for regularized regression. Challenge in establishing the asymptotic equivalence in Theorem \ref{thm:2} arises from that $\bbetahat\supbullet\subIPD$ and $\bbetahat\subadel\supbullet$  are not necessarily as sparse as the true coefficient $\bbeta_0\supbullet$. We overcome such challenge by comparing the two estimators in a more elaborative way as detailed in the proof. Moreover, Theorem \ref{thm:3} ensures the sparsistency of the SHIR under some commonly used conditions on deterministic design. Benefited from integrating the data, our approach requires less restrictive assumption on the minimum strength of the coefficients than the local estimation, under certain sparsity assumption. Such property is not readily available in the existing literature for distributed regression and the mixture penalty.

Our approach is also efficient both in computation and communication, as it only solves LASSO problem once in each local site without requiring the computation of $\bThetahat\supm$ or debiasing and only needs one around of communication. Consequently, since there is actually no debiasing procedure in our method, the SHIR cannot be directly used for hypothesis testing and confidence interval construction. Future work lies on developing statistical approaches for such purposes under DataSHIELD, high-dimensionality and heterogeneity.

{ 
For the choice of penalty, in the current paper we focus primarily on the mixture penalty, $\rho_2(\bbeta\supbullet)$ defined in (\ref{equ:pen}), because it can leverage the prior assumption on the similar support and magnitude of $\bbeta_0\supm$ across studies.
Nevertheless, other penalty functions such as those discussed in Section \ref{sec:int} can be incorporated into our framework provided that they can effectively leverage on the prior knowledge. Similar techniques used for deriving the theoretical results of SHIR with $\rho = \rho_2$ can be used for other penalty functions, with some technical details varying according to different choices on $\rho(\cdot)$. See Section A.4 of the Supplementary Material for further justifications.
}

We assume $n_m=\Theta(N/M)$ in Theorems \ref{thm:1}-\ref{thm:3} and allow $M$ to grow slowly. These conditions seem reasonable in our application but could be violated by some large scale meta-analysis of high dimensional data. The extensions to scenarios with larger $M$ is highly non-trivial and warrants further research. 
{ As a limitation of our theoretical analysis, our sparsity assumption in Theorem \ref{thm:2} may be somewhat strong for our real example. 
However, we shall clarify that, first, this assumption is comparable to the assumption for the estimation equivalence result in \cite{battey2018distributed}, and second, the consistency result in Theorem \ref{thm:1} holds with milder sparsity assumptions. On the other hand, it is of interests to see the possibilities of reducing the aggregation error and achieving the theoretical equivalence property with weaker sparsity assumption. One potential way is to use multiple rounds of communications such as \cite{fan2019communication}. Detailed analysis of this approach warrants future research.} %For the choice of penalty, we focus primarily on the mixture penalty in the current paper since it is a suitable choice for leveraging the prior assumption on the shared support and magnitude. However, other penalty functions underlying different sparse structures can be incorporated into our framework, such as the composite absolute penalties family  \citep{zhao2009composite} and the hierarchical LASSO \citep{cheng2015identification}. %Finally, our work is restricted to learning of the parameters for a parametric single index model with smooth and strongly convex loss (see Conditions \ref{cond:3} and \ref{cond:4}). We leave more explorations on the other models and loss functions beyond this to future research topics.
%And there is still lack of development in the methodology and asymptotic properties of the distributed version of many complicated statistical tools. 

\section{Supplementary Material}
In the supplement we provide some justifications for Conditions \ref{cond:1} and \ref{cond:7}, present detailed proofs of Theorems \ref{thm:1}--\ref{thm:3}, and outline theoretical analyses of SHIR for various penalty functions.

%This defines the bibliographies style. Search online for a list of available styles.

\bibliographystyle{apalike}
{\small \bibliography{library}

\newpage
\begin{figure}[H]
\centering
\caption{\label{fig:sim1a} The average absolute estimation error (AEE) of SHIR, Debias$\sublab$ and SMA relative to those of IPD under different $M$, $p$ and data generation mechanisms (i)--(iii) introduced in Section \ref{sec:simu}.} 
\includegraphics[width = 0.95\textwidth]{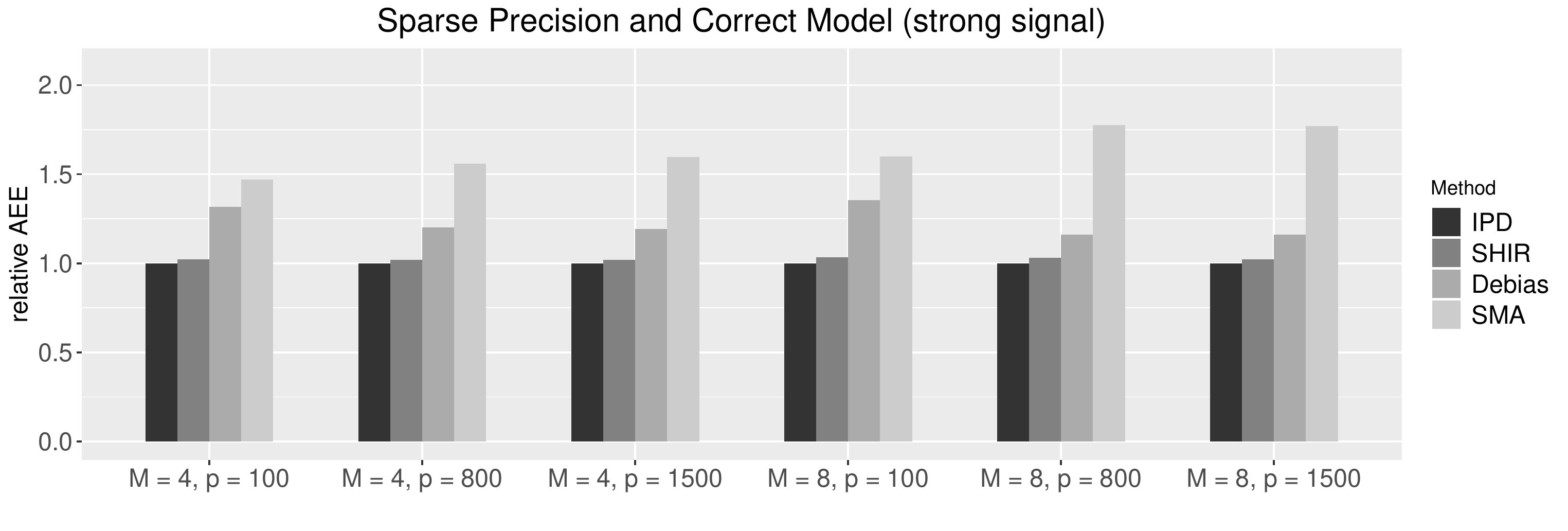}
\includegraphics[width = 0.95\textwidth]{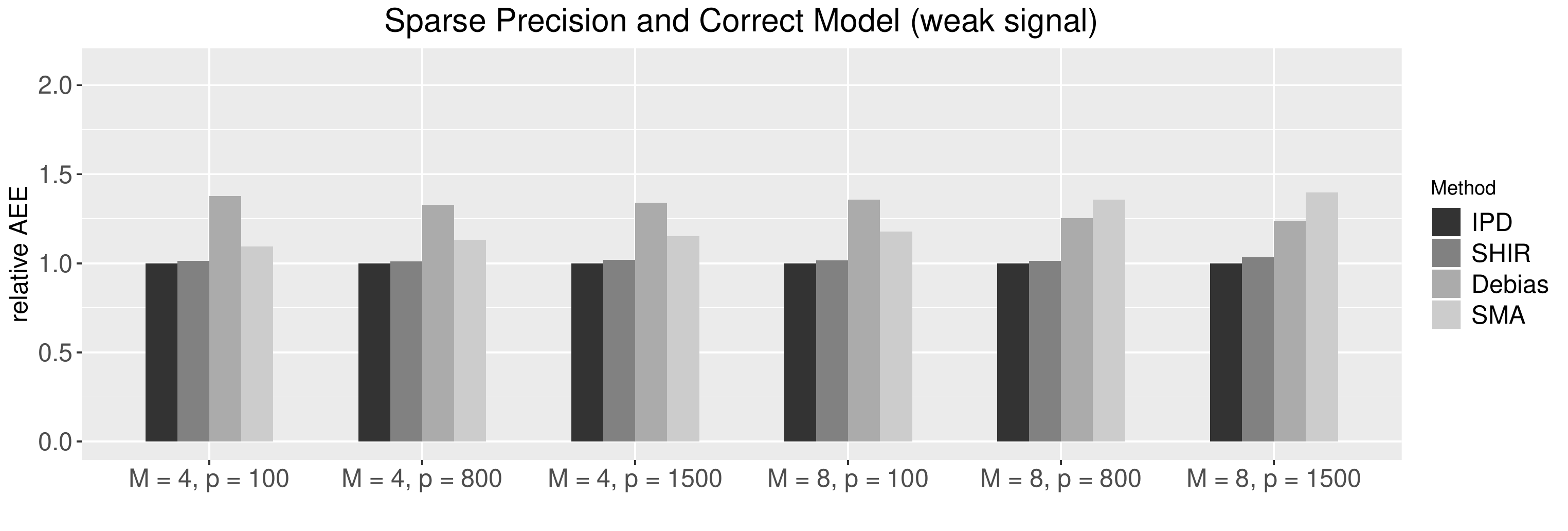}
\includegraphics[width = 0.95\textwidth]{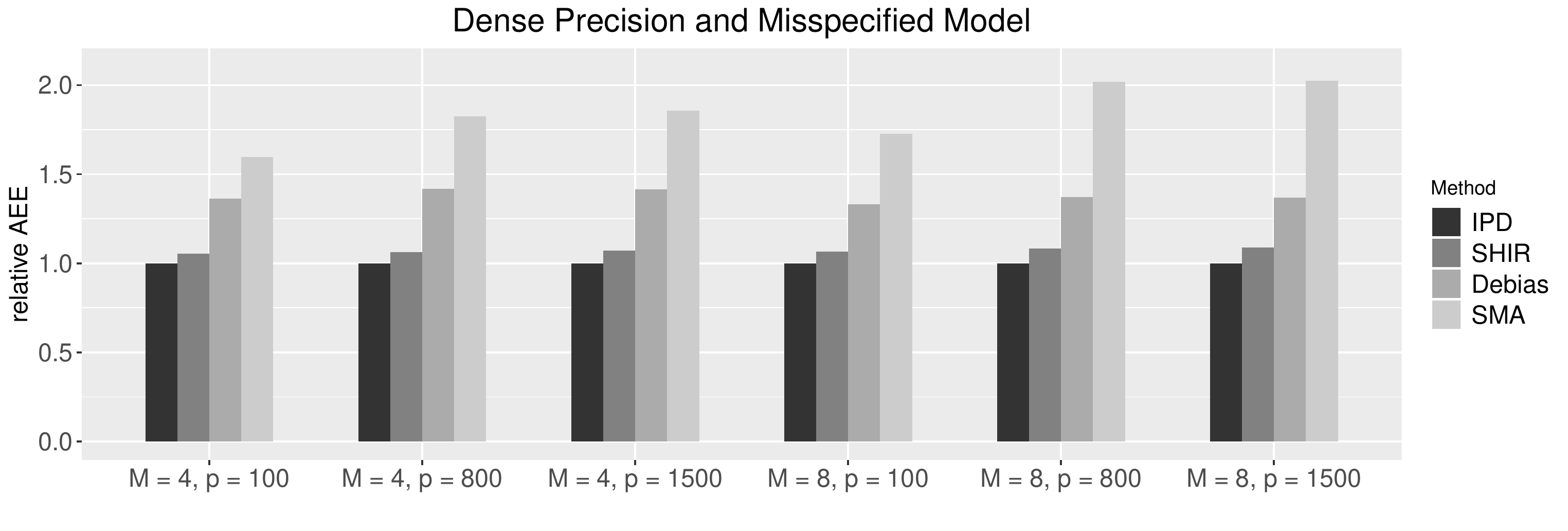}
\end{figure}

\newpage
\begin{figure}[H]
\centering
\caption{\label{fig:sim1b} The prediction error (PE) of SHIR, Debias$\sublab$ and SMA relative to those of IPD under different $M$, $p$ and data generation mechanisms (i)--(iii) introduced in Section \ref{sec:simu}.} 
\includegraphics[width = 0.95\textwidth]{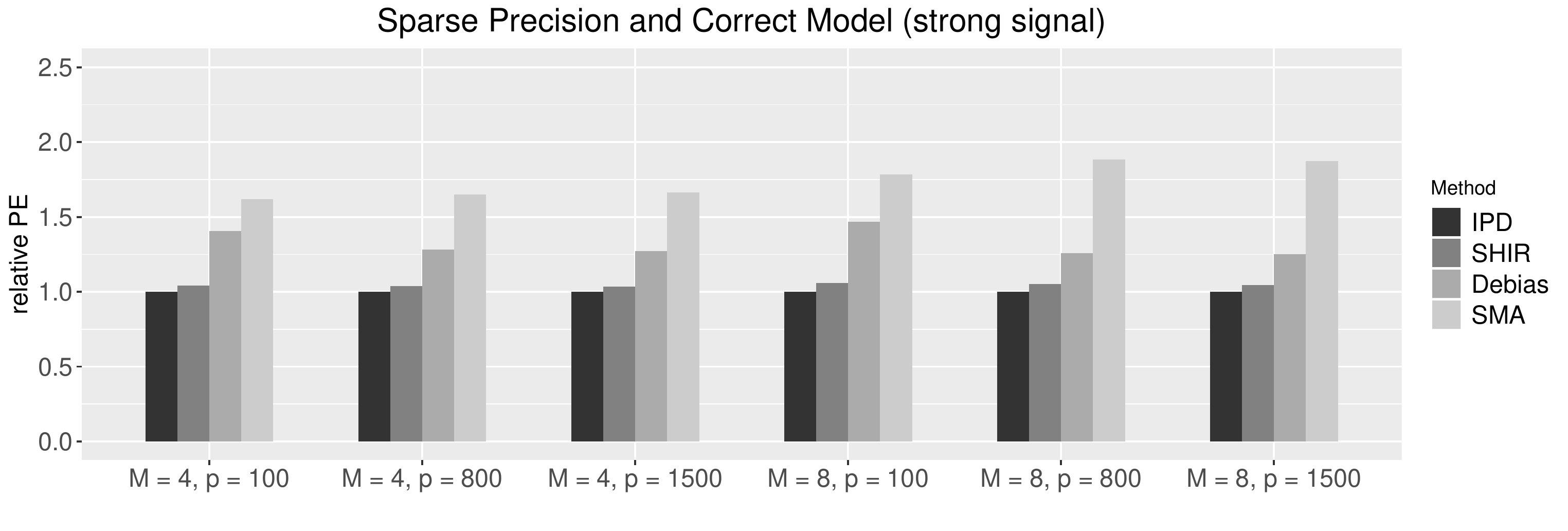}
\includegraphics[width = 0.95\textwidth]{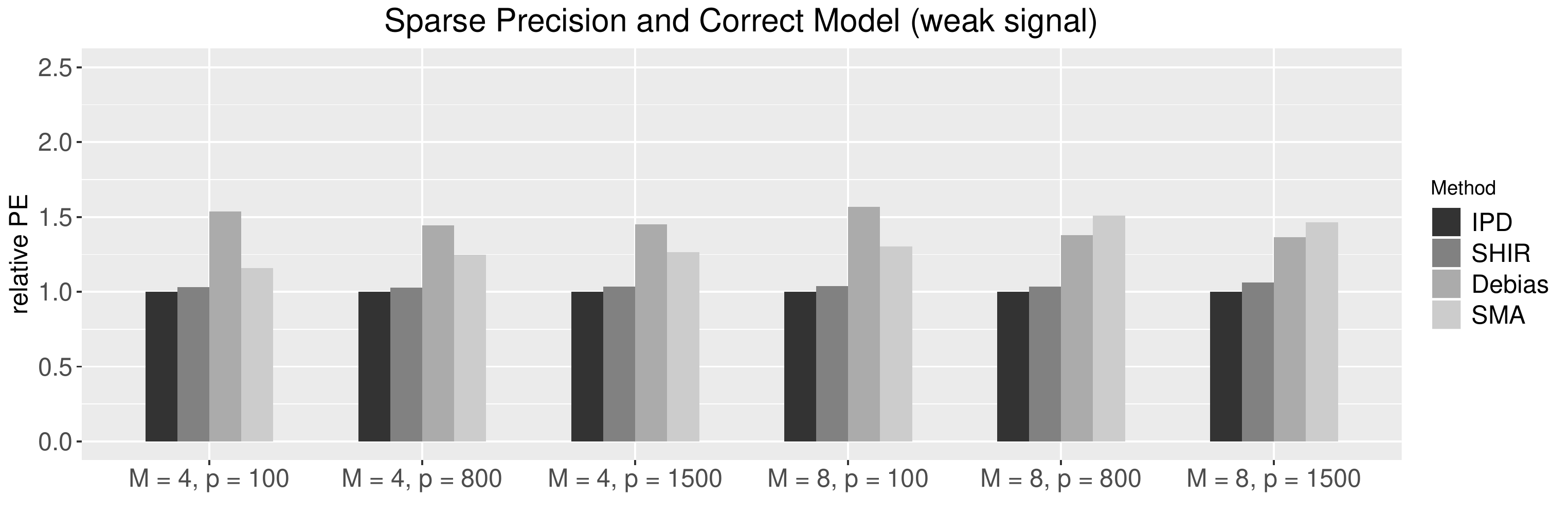}
\includegraphics[width = 0.95\textwidth]{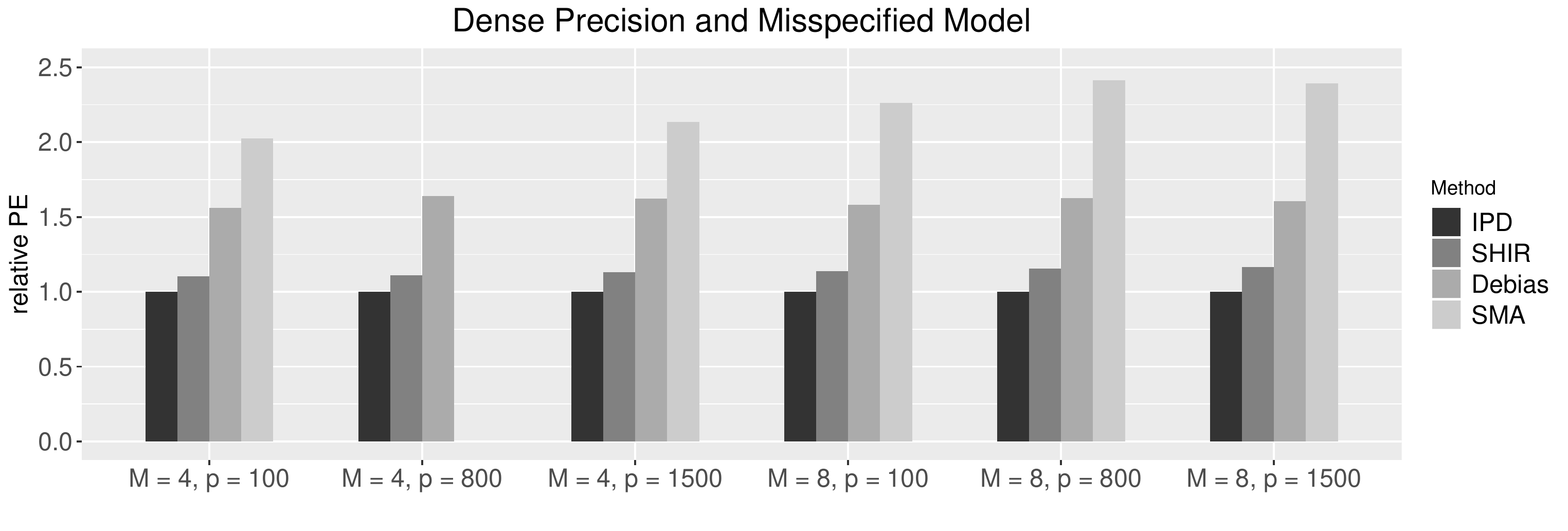}
\end{figure}

%\newpage
%\begin{figure}[H]
%\centering
%\caption{\label{fig:sim2a} The average true positive rate (TPR) and false discovery rate (FDR) on the homogeneous effect $\bmu$ of IPD, SHIR, Debias$\sublab$ and SMA, under different $M$, $p$ and the data generation mechanisms (i) (correct model with) strong signal and (ii) weak signal introduced in Section \ref{sec:simu}.}
%\includegraphics[width = 0.95\textwidth]{figure/tpr_homo_strong.pdf}
%\includegraphics[width = 0.95\textwidth]{figure/tpr_homo_weak.pdf}
%\includegraphics[width = 0.95\textwidth]{figure/fdr_homo_strong.pdf}
%\includegraphics[width = 0.95\textwidth]{figure/fdr_homo_weak.pdf}
%\end{figure}

%\newpage
%\begin{figure}[H]
%\centering
%\caption{ \label{fig:sim2b} The average true positive rate (TPR) and false discovery rate (FDR) on the heterogeneous effect $\balpha\supbullet$ of IPD, SHIR, Debias$\sublab$ and SMA, under different $M$, $p$ and the data generation mechanisms (i) (correct model with) strong signal and (ii) weak signal introduced in Section \ref{sec:simu}.}
%\includegraphics[width = 0.95\textwidth]{figure/tpr_heter_strong.pdf}
%\includegraphics[width = 0.95\textwidth]{figure/tpr_heter_weak.pdf}
%\includegraphics[width = 0.95\textwidth]{figure/fdr_heter_strong.pdf}
%\includegraphics[width = 0.95\textwidth]{figure/fdr_heter_weak.pdf}
%\end{figure}

\newpage
\begin{figure}[H]
\centering
\caption{\label{fig:sim2} The average true positive rate (TPR) and false discovery rate (FDR) on the original coefficients $\bbeta\supbullet$ of IPD, SHIR, Debias$\sublab$ and SMA, under different $M$, $p$ and the data generation mechanisms (i) (correct model with) strong signal and (ii) weak signal introduced in Section \ref{sec:simu}.}
\includegraphics[width = 0.95\textwidth]{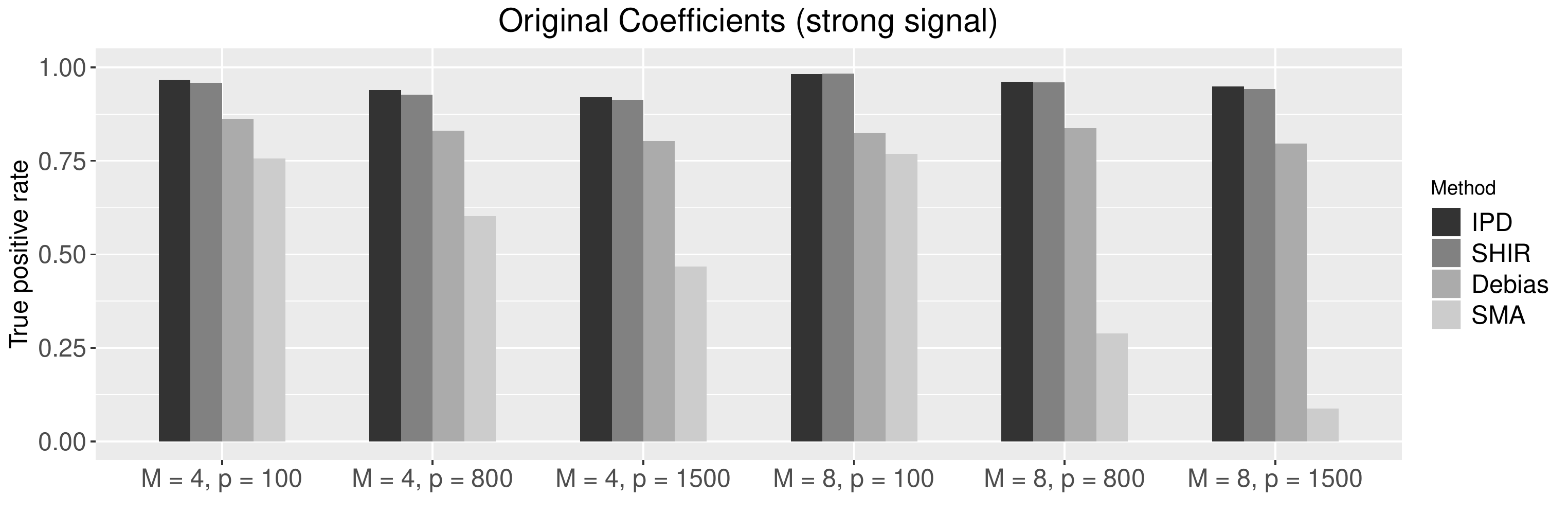}
\includegraphics[width = 0.95\textwidth]{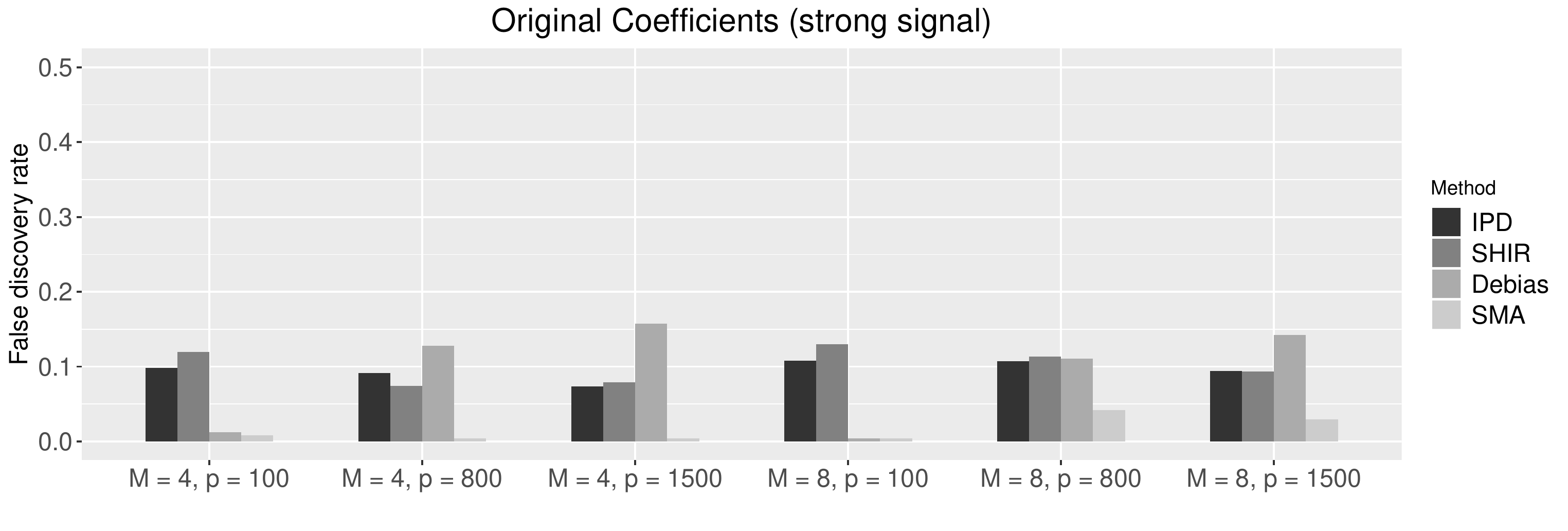}
\includegraphics[width = 0.95\textwidth]{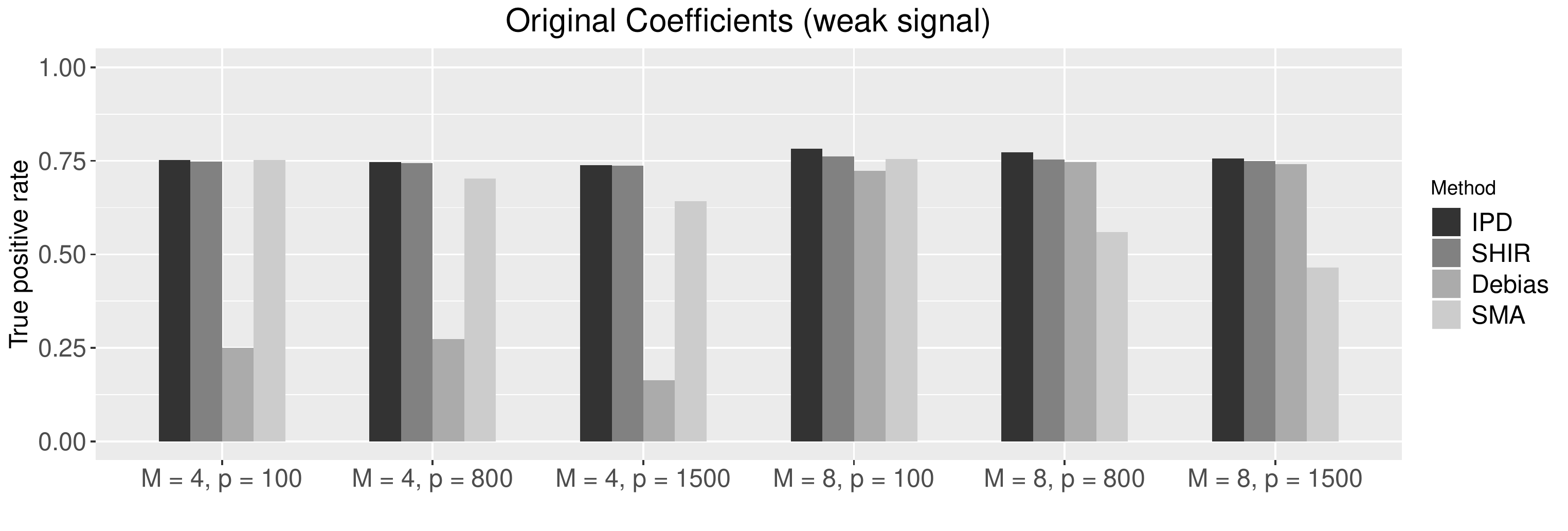}
\includegraphics[width = 0.95\textwidth]{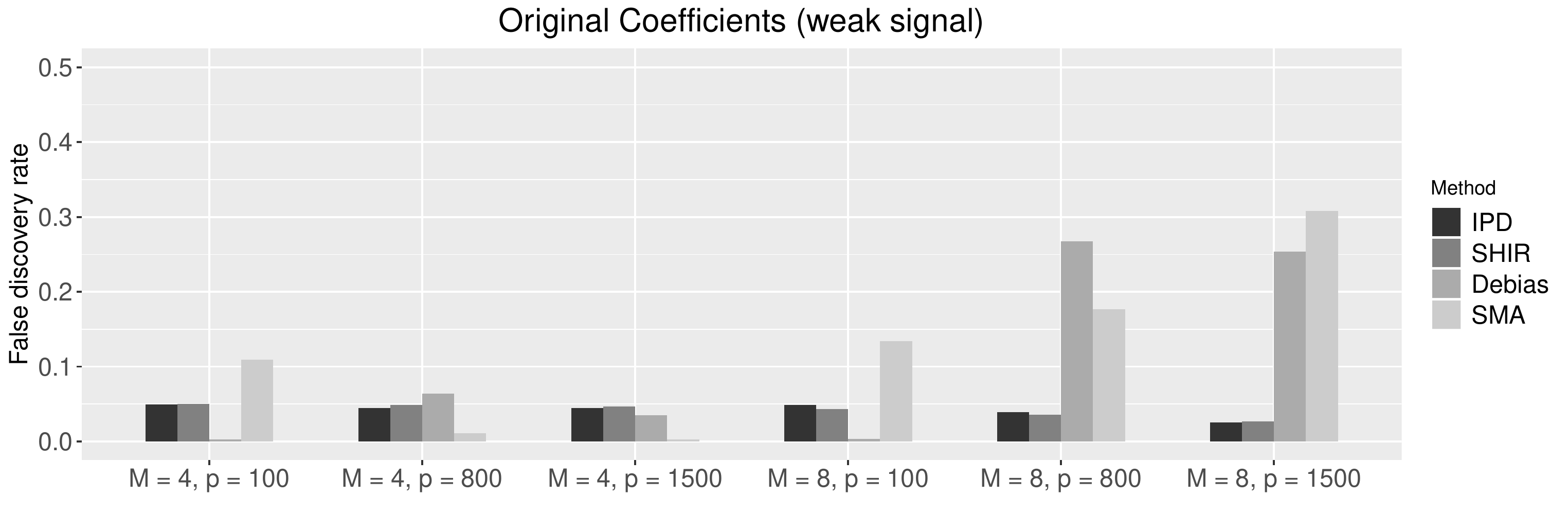}
\end{figure}

\newpage

\begin{table}[h]
\centering
\caption{Detected variables and magnitudes of their fitted coefficients for homogeneous effect $\mu$. A{\bf :}B denotes the interaction term of variables A and B. The $\log(x+1)$ transformation is taken on the count data and the covariates are normalized.}
~\\~
\begin{tabular}{|r|rrr|}
\hline
Variable                 & Debias$\sublab$ & SHIR  & SMA   \\ \hline
Prescription code of statin     & 0.14       & 0.07  & 0     \\ \hline
Age                & 0.09  & 0.26  & 0.28  \\ \hline
Procedure code for echo                & 0    & -0.10  & 0     \\ \hline
Total ICD counts           & -0.38  & -0.75 & 0     \\ \hline
NLP mention of CAD          & 0.97   & 1.34  & 0.81  \\ \hline
NLP mention of CAD procedure related concepts       & 0     & 0.02  & 0     \\ \hline
NLP mention of non-smoker      & -0.07  & -0.25 & -0.42 \\ \hline
ICD code for CAD               & 1.00    & 0.67  & 0.35  \\ \hline
CPT code for stent or CABG   & 0        & 0.05  & 0     \\ \hline
NLP mention of current-smoker    & 0         & -0.03 & 0     \\ \hline
Any NLP mention           & 0.06      & 0.05  & 0     \\ \hline
ICD code for CAD{\bf :}Procedure code for echo  & 0          & -0.04 & 0     \\ \hline
NLP mention of CAD:NLP mention of possible-smoker & 0         & -0.02 & 0     \\ \hline
Oncall:NLP mention of non-smoker   & 0.09       & 0     & 0     \\ \hline
Indication for NLP mention of non-smoker    & -0.53      & 0     & 0     \\ \hline
\end{tabular}
\label{tab:var}
\end{table}

\newpage

\begin{figure}[H]
\centering
\caption{\label{fig:1} The mean and 95\% bootstrap confidence interval of AUC, Brier Score, $F_{5\%}$ and $F_{10\%}$ of Debias$\sublab$,  Local, SHIR and SMA on the validation data from the four studies. }
%\tcomm{put methods together then by disease, simplify titles to AUC, F1 Score etc. show point estimate on validation data not median of bootstrap}
  \begin{minipage}[h]{7in}
  \centering
  \mbox{
    \includegraphics[width=3.5in]{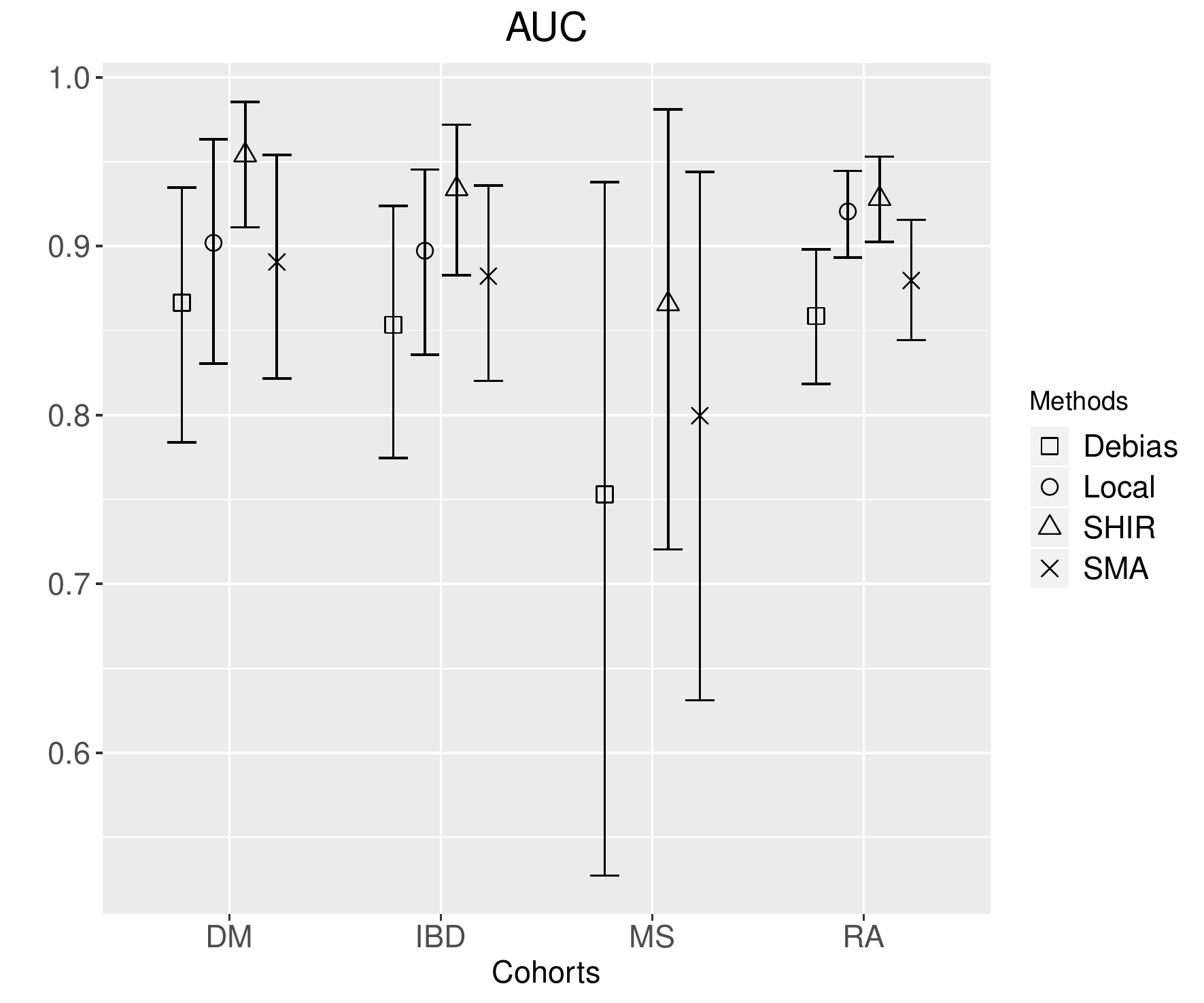}
    \includegraphics[width=3.5in]{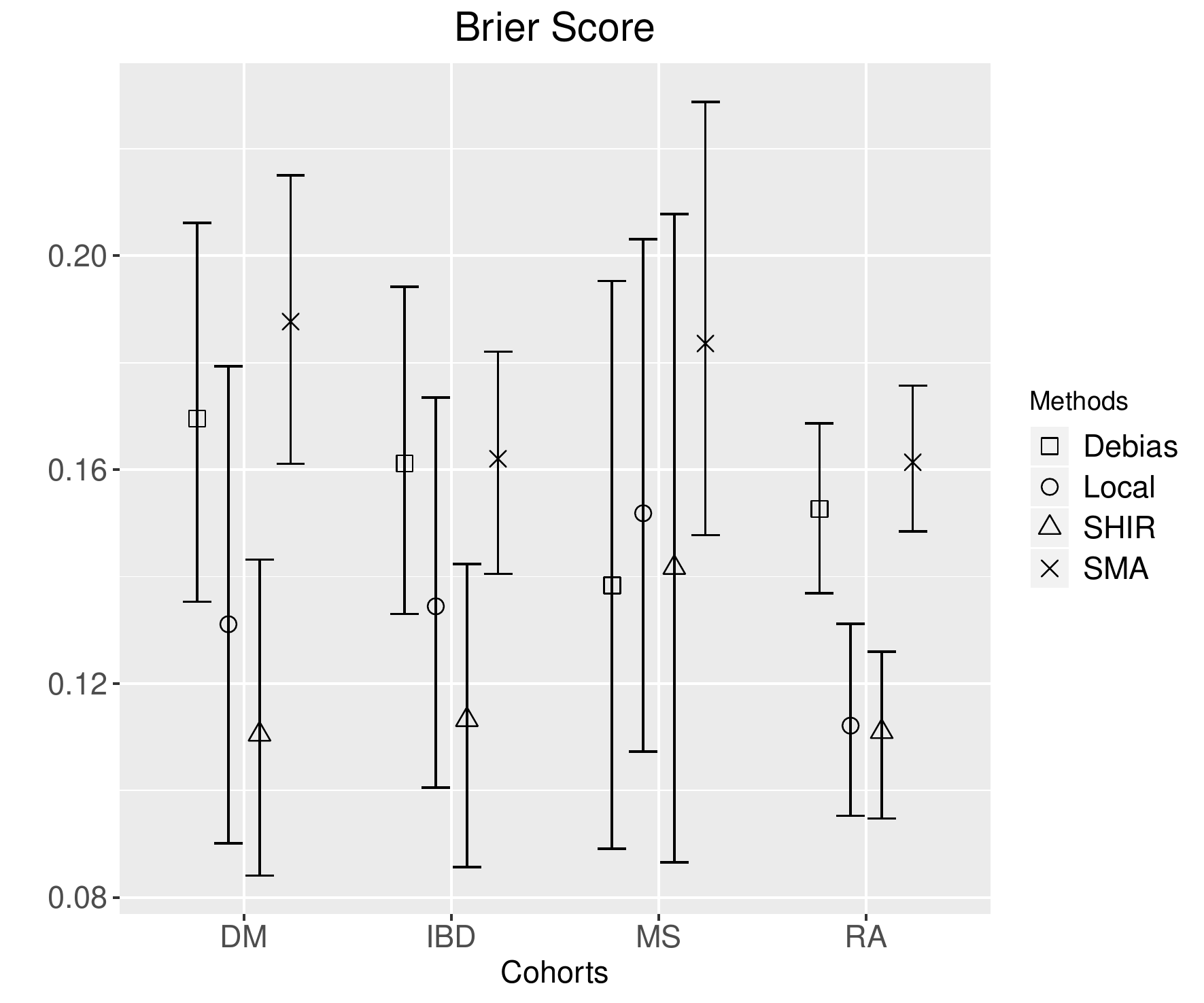}
    }
    \mbox{
    \includegraphics[width=3.5in]{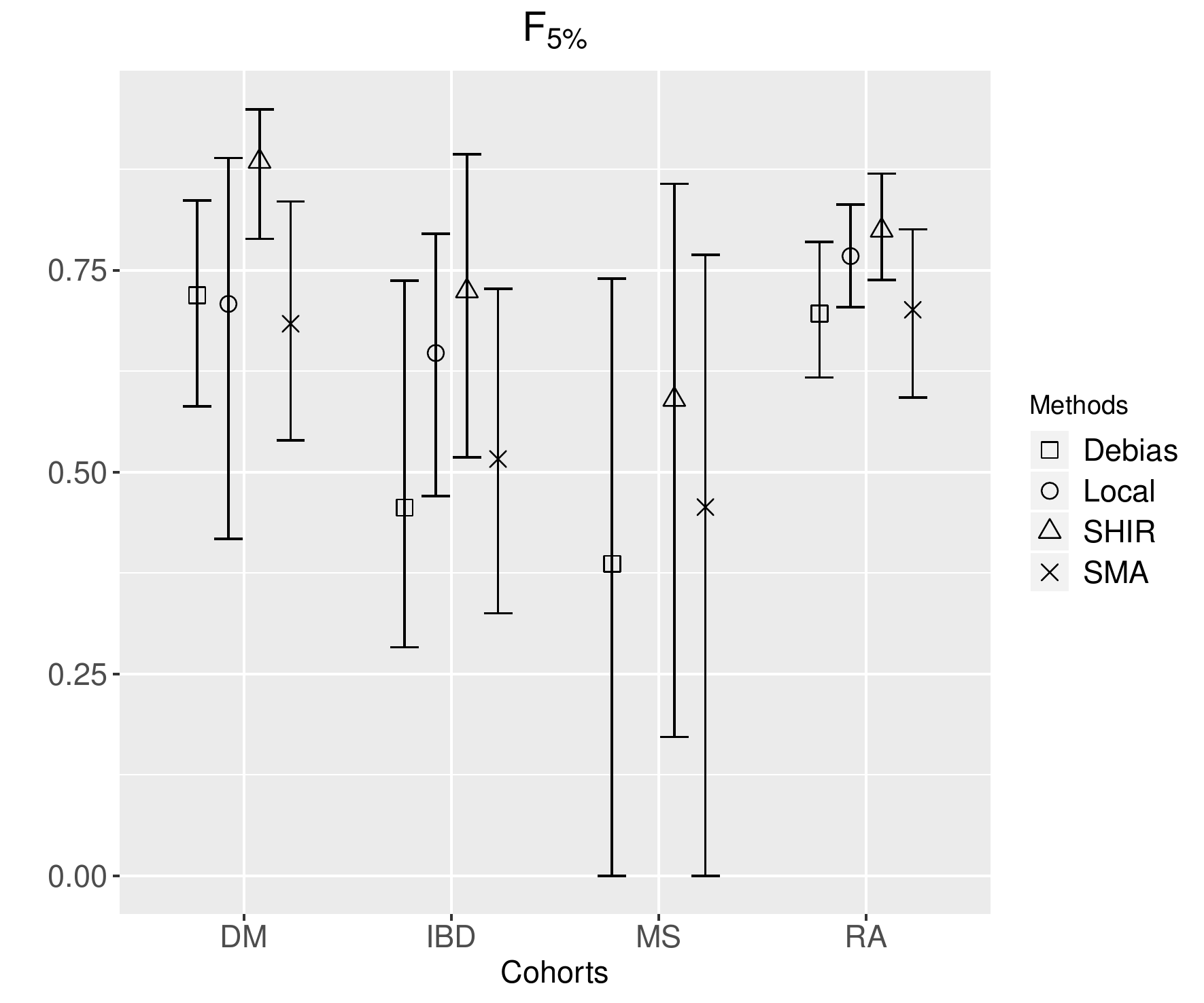}
    \includegraphics[width=3.5in]{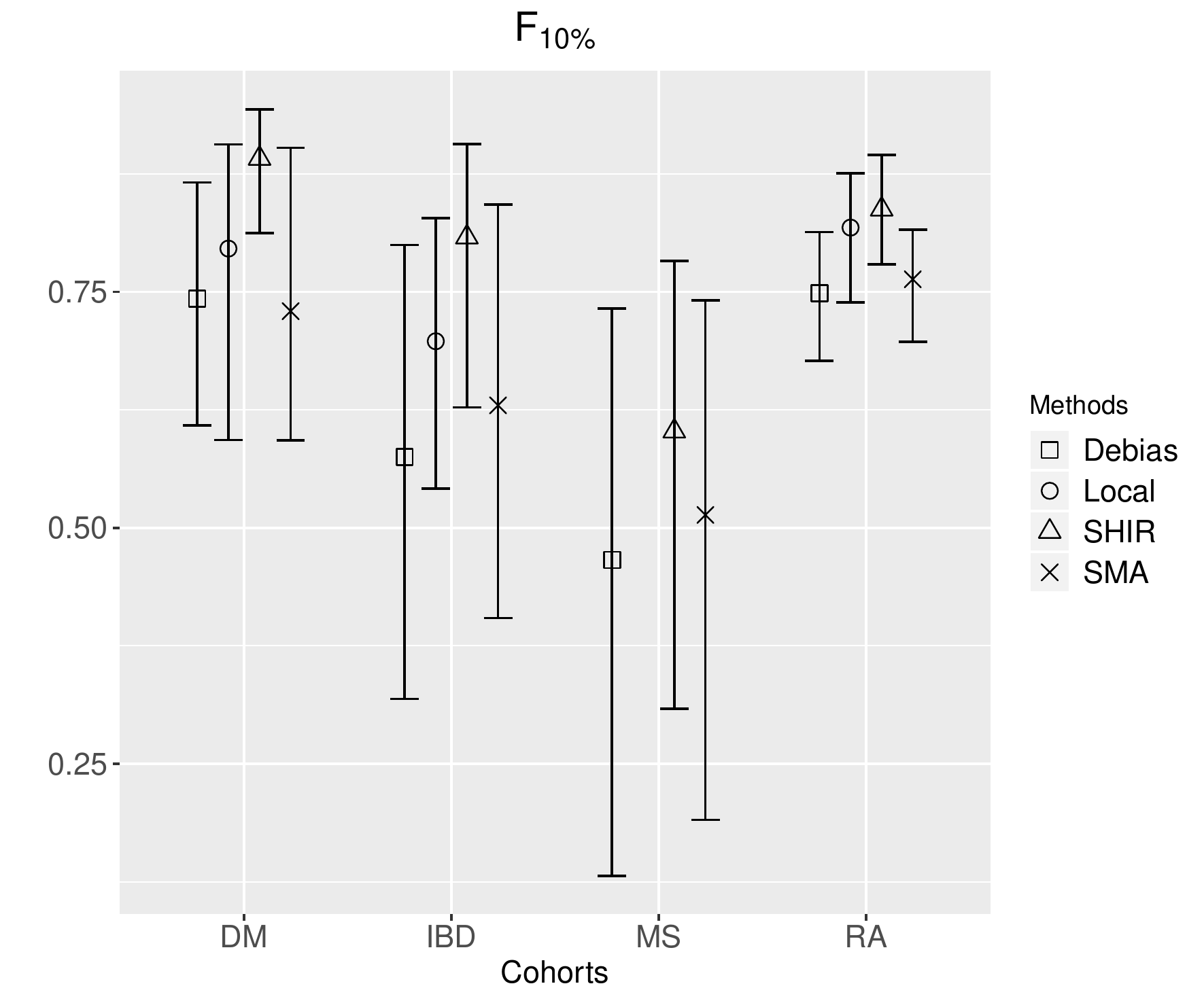}
    }

  \end{minipage} 
\end{figure}

\end{document}